\documentclass[aps,prd,preprint,superscriptaddress,,nofootinbib,amsmath,amssymb]{revtex4}
\usepackage{graphicx}
\usepackage{dcolumn}
\usepackage{bm}
\usepackage{amsmath}
\usepackage{amsfonts}
\usepackage{amssymb}
\usepackage{appendix}
\usepackage{pdfpages}
\usepackage{graphicx}
\usepackage{wrapfig}
\usepackage{multirow}
\usepackage{mathrsfs}
\usepackage{epstopdf}
\usepackage{slashed}
\usepackage{soul}
\usepackage{mathtools}
\usepackage{floatrow}
\usepackage{float}
\usepackage{csquotes}
\usepackage{wrapfig}
\usepackage{graphicx}
\usepackage[dvipdfm]{hyperref}
\usepackage{epsfig}
\usepackage{hyperref}
\usepackage[utf8]{inputenc}
\usepackage{epsfig}
\usepackage{dcolumn}
\usepackage{morefloats}

\def\be{\begin{equation}}
\def\ee{\end{equation}}
\def\bea{\begin{eqnarray}}
\def\eea{\end{eqnarray}}
\def\gsim{\ \rlap{\raise 2pt\hbox{$>$}}{\lower 2pt \hbox{$\sim$}}\ }
\def\lsim{\ \rlap{\raise 2pt\hbox{$<$}}{\lower 2pt \hbox{$\sim$}}\ }
\def\dslash{\kern-4pt \not{\hbox{\kern-2pt $\partial$}}}
\def\pslash{\not{\hbox{\kern-2pt p}}}

\newcommand{\dcp}{\delta_{CP}}
\newcommand{\nova}{NO$\nu$A}

\begin{document}
\DeclareGraphicsExtensions{.eps,.ps}


\title{\boldmath New look at the degeneracies in the neutrino oscillation 
parameters, and their resolution by T2K, \nova\  and ICAL }



\author{Monojit Ghosh}
\email[Email Address: ]{monojit@prl.res.in}
\affiliation{
Physical Research Laboratory, Navrangpura,
Ahmedabad 380 009, India}
 
\author{Pomita Ghoshal}
\email[Email Address: ]{pomita.ghoshal@gmail.com}
\affiliation{
Department of Physics, LNM Institute of Information Technology (LNMIIT),
Rupa-ki-Nangal, post-Sumel, via-Jamdoli, Jaipur-302 031, Rajasthan, India}
 
\author{Srubabati Goswami}
\email[Email Address: ]{sruba@prl.res.in}
\affiliation{
Physical Research Laboratory, Navrangpura,
Ahmedabad 380 009, India}

\author{Newton Nath}
\email[Email Address: ]{newton@prl.res.in}
\affiliation{
Physical Research Laboratory, Navrangpura,
Ahmedabad 380 009, India}
\affiliation{Indian Institute of Technology, Gandhinagar, Ahmedabad--382424, India}

\author{Sushant K. Raut}
\email[Email Address: ]{raut@kth.se}
\affiliation{
Department of Theoretical Physics, School of Engineering Sciences,
KTH Royal Institute of Technology – AlbaNova University Center,
Roslagstullsbacken 21, 106 91 Stockholm, Sweden}

\begin{abstract}
The three major unknown neutrino oscillation parameters at the present juncture are the mass hierarchy, the octant of the mixing angle $\theta_{23}$ and the CP phase $\dcp$. It is well known that the presence of hierarchy$-\dcp$ and octant 
degeneracies affects  the unambiguous determination of these  parameters. 
In this paper we show that a comprehensive way to study the remaining parameter degeneracies is in the form of a generalized hierarchy$- \theta_{23} - \dcp$ degeneracy. 
This is best depicted as contours in the test ($\theta_{23} - \dcp$) plane for 
different representative true values of parameters. We  show that  
the wrong-hierarchy and/or wrong-octant solutions  can be further classified 
into  eight different  solutions 
depending on whether they 
occur  
with the wrong or right value of $\dcp$.  These eight solutions are different 
from the original eightfold degenerate solutions
and can exist, in principle, even if $\theta_{13}$ is known.  
These multiple solutions, apart from  affecting the determination of  the true hierarchy and octant, also   affect the accurate estimation of $\dcp$. We identify  which of these eight different degenerate solutions  can occur in the test ($\theta_{23} - \dcp$) parameter space, taking the long-baseline experiment \nova\ running in the neutrino 
mode as an example. The inclusion of the \nova\ antineutrino run removes the wrong-octant solutions  appearing with both right and wrong-hierarchy. Adding T2K data to this resolves the wrong hierarchy -- right octant solutions to a large extent. The remaining wrong-hierarchy solutions can be removed by combining \nova\ + T2K  with atmospheric neutrino data. We demonstrate this using ICAL@INO as the prototype atmospheric neutrino 
detector.  We find that the degeneracies can be resolved at the  $2\sigma$ level by the combined data set, for the true parameter space considered in the study. 
\end{abstract}
\maketitle

\section{Introduction}
The standard three-flavor neutrino oscillation probability is described by 
six parameters, namely three mixing angles -- $\theta_{12}$, $\theta_{23}$, $\theta_{13}$, 
two mass squared differences -- $\Delta m^2_{31},~\Delta m^2_{21}$ ($\Delta m^2_{ij} = m_i^2 - m_j^2$) and the Dirac CP phase $\dcp$. The neutrino oscillation data from solar, atmospheric, reactor and accelerator experiments have so far given information about each of these oscillation parameters except $\delta_{CP}$ \cite{global_nufit,global_fogli,global_valle}. At present, the unknowns in neutrino oscillation physics are (i) the sign of $\Delta m^2_{31}$ [$\Delta m^2_{31} >0$  known as 
normal hierarchy (NH) and  $\Delta m^2_{31}<0$ known as inverted hierarchy (IH)], (ii) the octant of $\theta_{23}$ [$\theta_{23} > 45^\circ$ known as Higher Octant (HO) and  $\theta_{23} < 45^\circ$ known as  Lower Octant (LO)]. (iii) the CP phase $\dcp$; any value of this parameter other than $0^\circ$ and $ \pm 180^\circ$ would signal CP violation in the lepton sector. In this case, it is often useful to talk in terms of the lower half-plane (LHP) with $-180^\circ < \dcp < 0^\circ$ and upper half-plane (UHP) with 
$ 0^\circ < \dcp < 180^\circ$.  

The appearance channel $P_{\mu e}$ often known as the \enquote{golden channel}  
can measure all the three unknown parameters described above\footnote{Originally, $P_{\mu e}$ was termed as the golden channel because of its sensitivity to $\theta_{13}$, hierarchy and $\dcp$.}. However, the measurement is complicated by the fact
that different sets of values of parameters can give the same oscillation probability. 
This gives rise to  degeneracies that render an unambiguous determination of 
true parameters difficult. It was discussed in Ref.~\cite{barger} that there can be eight-fold degeneracies in neutrino oscillation probabilities which are 
(a) the intrinsic or $\theta_{13} -\dcp$ degeneracy ~\cite{intrinsic}, (b) the hierarchy-$\dcp$ degeneracy ~\cite{Minakata:2001qm} and (c) the intrinsic octant degeneracy ~\cite{lisi}. The intrinsic degeneracy refers to clone solutions occurring due to a different $\theta_{13}$ and $\dcp$ value. This degeneracy can be removed
to a large extent by using spectral information \cite{twobase2}.
Moreover, the current precision determination of $\theta_{13}$  ~\cite{t2k_t13_jun2011, dchooz_1406,dayabay_latest,reno_t13} has removed 
this degeneracy to a great extent. The hierarchy-$\dcp$
degeneracy leads to wrong-hierarchy solutions occurring for a different 
value of $\dcp$ other than the  true value. The intrinsic octant degeneracy 
refers to duplicate solutions occurring for $\theta_{23}$ and $\pi/2-\theta_{23}$. 

Many papers have discussed possibilities of the resolution of these  
degeneracies by using different detectors in the same experiment  
\cite{Donini:2002rm,degeneracy1,twobase5}. The synergistic combination of 
data from different experiments was also discussed as an effective means 
of removing such degeneracies by virtue of the fact that the oscillation probabilities offer different combinations of parameters at varying
baselines and energies \cite{Narayan:1999ck,twobase1,twobase2,twobase3,twobase6,synergynt,
Huber:2003pm,menaparke,Mena:2005ek}. In particular, the synergy between 
long-baseline (LBL) experiments \nova\ and T2K in resolving the 
hierarchy-$\dcp$ degeneracy has been discussed recently in 
Refs.\cite{Minakata:2003wq,suprabh_t2knova,sanjib_glade,gainfracs}. 

It has been shown in Refs.\cite{Huber:2003pm,minakata,minakata2} that 
a precise measurement of the mixing angle $\theta_{13}$ is helpful 
for the removal of octant degeneracy.
Octant sensitivity in the T2K and NO$ \nu $A  experiments has been studied
recently  in Refs.\cite{sushant_2012,usoctant} in view of the measurement of a non zero $\theta_{13}$.  
The octant degeneracy is different for neutrinos and antineutrinos and hence a combination of these two data sets can be conducive for the removal of this degeneracy for most values of $\dcp$ \cite{suprabhoctant,minakata_cp,Ghosh:2014zea}. 

Since atmospheric neutrino baselines experience strong Earth matter effects, 
these effectively remove the overlap between right and wrong-hierarchy solutions \cite{atmoshier1,atmoshier2,atmoshier3,atmoshier6}. In particular, atmospheric neutrino experiments capable of distinguishing neutrinos and antineutrinos can be very useful in resolving degeneracies related to the mass hierarchy 
\cite{atmoshier4,atmoshier5,atmoshier6,atmoshier7,ushier,Samanta:2006sj, 
ino3d,gct,Ghosh:2013mga,lar1,Barger:2012fx}. The octant sensitivity of the
atmospheric neutrinos comes from both the appearance \cite{dev_ggms} and 
disappearance channels \cite{Choubey:2005zy}, and also benefits from 
significant matter effects, especially facilitated by the large value of 
$\theta_{13}$ measured by reactor experiments. Atmospheric neutrinos also 
provide a synergy with LBL experiments in terms of probability behavior with respect to
parameters, so that the combination of atmospheric neutrino data with LBL data exhibits
reduced effect of the hierarchy and octant degeneracies 
\cite{Huber:2005ep,schwetzblennow,gct,usoctant,Choubey:2013xqa}.

Recently, it has been realized that for the appearance channel, the octant 
degeneracy  can be generalized to the octant-$\dcp$ degeneracy corresponding to any value of $\theta_{23}$ in the opposite octant \cite{usoctant, coloma}. A continuous generalized  degeneracy in the three-dimensional $\theta_{23}-\theta_{13} -\dcp$ plane has been studied in Ref.\cite{coloma}. In this work, we show that with the high precision measurement of 
$\theta_{13}$ by reactor experiments, the degeneracies can be discussed in an integrated manner  in terms of a  generalized hierarchy-$\theta_{23}$ - $\dcp$ degeneracy.    
A good way to visualize the different degenerate solutions is in terms of contours in the test ($\theta_{23} - \dcp$) plane for different choices of true values of parameters\footnote{Note that prior to the discovery of a nonzero value of $\theta_{13}$, the degeneracies were studied mainly in $\theta_{13} - \dcp$ plane. }. These plots also give an indication regarding the precision of the parameters $\dcp$ and $\theta_{23}$. 
Although hierarchy degeneracy is discrete, the $\theta_{23}-\dcp$ degeneracy is continuous for the appearance channel probability $P_{\mu e}$. Inclusion of the information from the disappearance channel $P_{\mu \mu}$ restricts $\theta_{23}$ and discrete degenerate solutions are generated. We classify,  for the first time the wrong-hierarchy and wrong-octant solutions with respect to right or wrong $\dcp$ values. This also allows us to understand how the hierarchy and octant degeneracies can affect the precision in $\dcp$. We observe that since the wrong-hierarchy and wrong-octant solutions can occur for wrong values of $\dcp$ as well,  there can exist, in principle, a total of eight degenerate solutions corresponding to different combinations of hierarchy, octant and $\dcp$. This is summarized in Table~\ref{degneracy_table}{\footnote{ It is to be noted in this connection that if the $\dcp$ precision is not good then there can be continuous regions
connecting right and wrong $\dcp$ solutions and hence it may not always 
be possible to identify discrete wrong $\dcp$ solutions.}}.  Note that these solutions are different from the eight-fold degenerate solutions that have been discussed in the literature.  To the best of our knowledge the parameter degeneracies have not been studied in this generalized form in the literature prior to this. We identify which degenerate solutions among the eight possibilities listed in Table~\ref{degneracy_table} exist in the neutrino oscillation probabilities for typical baselines and energies corresponding to the LBL experiments T2K and \nova. 

For representative true values of 
these parameters, we demonstrate to what extent the degenerate solutions can 
be removed by \nova, \nova+T2K and \nova+T2K+ICAL. Note that although the combined capability of \nova, T2K and ICAL in hierarchy octant and $\dcp$ determination have been investigated, a comprehensive study for removal of degeneracies using these three facilities together has not been done before. 
  

The paper is organized as follows. In Sec. II, we give the experimental details of the LBL and atmospheric neutrino experiments being considered. In Sec. III, first we summarize the parameter degeneracies and identify degenerate solutions at the level of 
neutrino oscillation probabilities. Then, we show their occurrence at the event level considering \nova\ and discuss the resolution of the different kinds of degeneracies by combinations of the given experiments.  We also present the precision of the parameters $\theta_{23}$ and $\dcp$ from the combined analysis with \nova+T2K+ICAL data. The conclusions are presented in Sec. IV. The Appendix A outlines the synergy between the disappearance and appearance channels and the role of antineutrinos.

\begin{table}[h]
\centering
\begin{tabular}{|c|c|}
\hline
Solution with  & Solution with  \\
right $\delta_{CP} $ &  wrong $\delta_{CP} $  \\
\hline
I. RH-RO-R$\delta_{CP} $ & V. WH-WO-W$\delta_{CP} $ \\
II. RH-WO-R$\delta_{CP} $ & VI. RH-RO-W$\delta_{CP} $ \\
III. WH-RO-R$\delta_{CP} $ & VII. RH-WO-W$\delta_{CP} $ \\
IV. WH-WO-R$\delta_{CP} $ & VIII. WH-RO-W$\delta_{CP} $ \\
\hline
\end{tabular}
\vspace{3mm}
\caption{\footnotesize Various possibilities of degeneracy in 
the probability $P_{\mu e}$.
Here, R=right, W=wrong, H=hierarchy and O=octant.}
\label{degneracy_table}
\end{table}

\section{ Experimental Details }
We use the GLoBES package~\cite{globes1,globes2} (along with the required auxiliary files~\cite{messier_xsec,paschos_xsec}) to simulate the data of the two long-baseline neutrino oscillation experiments T2K (Tokai to Kamioka, Japan) and  \nova\ (NuMI Off-Axis $ \nu_{e} $ Appearance, Fermilab). The source to detector distance, $L$ for T2K is 295 km. In the T2K experiment \cite{t2k}, muon neutrinos are directed from J-PARC, making an off-axis angle of $2.5^\circ$ toward the Super-Kamiokande detector, which is a Water \v{C}erenkov detector of mass 22.5 kt. T2K has been proposed to run for a total luminosity of $7.8 \times 10^{21}$ protons on target (POT), and it has already collected $10\%$ of the total data in the neutrino mode. At present it is running in the antineutrino mode, and the  first results have been reported ~\cite{t2k_antinu_website}. The recently operational \nova experiment is also sending muon neutrinos through two detectors, one at Fermilab  (the near detector) and one in northern Minnesota (the far detector), making an off-axis angle of $0.8^\circ  $ and traveling a distance of 810 km to reach the far detector, which is a 14 kt totally active scintillator detector (TASD). The beam power of \nova\ is planned to be 700 kW which corresponds to $ 6 \times 10^{20} $ POT/year which will run for $ 3(\nu) + 3(\overline{\nu}) $ years. In our simulation of \nova\ data, we considered the reoptimized \nova\ set up from Refs.\cite{sanjib_glade,Kyoto2012nova}.

For the simulation of an atmospheric neutrino experiment, we consider a magnetized iron calorimeter detector (ICAL)  planned by the India-based Neutrino Observatory (INO),
 the primary goal of which is to study atmospheric muon neutrinos. ICAL@INO has  an advantage over other detectors because it has a magnetic field which allows charge discrimination, thus providing the facility to distinguish between $ \mu^{+} $ and $ \mu^{-} $.  Here, we consider the 50 kt detector for ICAL@INO with a runtime of 10 years. In our numerical analysis we have used constant detector angular and energy resolutions 
of 10$ ^\circ $ and 10\% respectively  and 85\% overall efficiency.  The muon resolutions from INO simulations can be found in Ref.\cite{ino2d}. We have checked that the constant resolutions used in this paper give results similar to those obtained using resolutions from INO simulation codes \cite{gct,ino3d}. In our analysis of atmospheric neutrinos, we used the Gaussian formula to compute the statistical $\chi^{2}$. Systematic errors are taken into account using the method of pulls~\cite{pulls_gg,pull_lisi} as outlined in Ref.\cite{ushier}. We have added a  $5\%$ prior on $ \sin^22\theta_{13}$. 

We use the following representative values for the oscillation 
parameters in our numerical simulation as given
in Table \ref{param_values} 
Refs.\cite{global_nufit,global_fogli,global_valle}.
\begin{table}[H]
\begin{center}
\begin{tabular}{|c|c|c|}
\hline 
Oscillation parameters & True value & Test value \\ 
\hline
$ \sin^{2}2 \theta_{13} $ & 0.1 & 0.085 -- 0.115 \\ 
$ \sin^{2}\theta_{12} $ & 0.31 & -- \\
$ \theta_{23} $ & LO=$39^\circ,42^\circ$, HO=$48^\circ, 51^\circ$ & $35^\circ-55^\circ$ \\
$\Delta m^2_{21} $ & 7.60 $ \times 10^{-5} eV^{2} $ & -- \\
$\Delta m^2_{31} $ & 2.40 $ \times 10^{-3}eV^{2}  $ & (2.15 -- 2.65) $ \times 10^{-3} eV^{2} $ \\
$ \delta_{CP} $ & $ 90^\circ $, $ 0^\circ $,$ - 90^\circ $ & $ -180^\circ $ to $ +180^\circ $ \\
\hline
\end{tabular}
\caption{\footnotesize Values of neutrino oscillation parameters used in our 
simulations. Here, the second column gives the true values of the parameters, 
and the third column represents the parameter range over which we have 
marginalized the test values.}
\label{param_values}
\end{center}
\end{table}
\section{Identifying Degeneracies  in neutrino oscillation 
parameters and their resolution  }
For the baselines relevant for the experiments \nova\ and T2K, the Earth matter density is in the range (2.3 -- 3.0 g/cc), well below the matter resonance. Therefore oscillation probabilities computed assuming constant matter density can be used for these experiments.
Such probabilities calculated using perturbative expansion of the small  leptonic mixing angle $ \theta_{13} $ (in terms of $ s_{13} $) and the mass hierarchy parameters $ \alpha$  $(\equiv \Delta m^{2}_{21}/ \Delta m^{2}_{31}) $ are given as follows
\cite{akhmedov,cervera,freund},
\begin{align}
P_{\mu\mu} &= 1 - \sin^2 2\theta_{23} \sin^2\Delta + \mathcal{O}( \alpha , s_{13}) \\
\label{p_mu_mu}
P_{\mu e} &= 4 s^{2}_{13}s^{2}_{23}\frac{\sin^{2} (A-1)\Delta}{(A-1)^2}  
    + \alpha^{2} \cos^{2}\theta_{23} \sin^{2}2 \theta_{12} \frac{\sin^{2} A\Delta}{A^2} \\ \nonumber
     & +\alpha s_{13} \sin 2\theta_{12}  \sin 2\theta_{23}\cos(\Delta+\delta_{cp}) \frac{\sin (A-1)\Delta}{(A-1)}\frac{\sin A\Delta}{A} \\ \nonumber
\label{p_mu_e}
\end{align}
where $ s_{ij}(c_{ij})=\sin \theta_{ij}(\cos \theta_{ij}) $ for 
$ j>i $ ($ i,j = 1,2,3 $). We use the following notation:
$ \Delta \equiv  \Delta m^{2}_{31} L / 4 E $ , 
$ A \equiv 2E V/ \Delta m^{2}_{31} = VL/2 \Delta $, 
where $ V =  \sqrt{2} G_F n_e$    
is the Wolfenstein matter term.
The antineutrino oscillation probability can be obtained by replacing 
$\delta_{CP} \rightarrow - \delta_{CP}  $ and $ V \rightarrow  - V $.  
Hence, in the neutrino oscillation probability, $A$ is positive for 
NH and negative for IH, and for antineutrinos, the sign of $A$ gets reversed.
We observe the following salient features from the probability formulas:
\begin{itemize}
\item The leading order term in the muon neutrino survival probability 
$ P_{\mu\mu} $, also known as disappearance channel, is proportional 
to $ \sin^2 2\theta_{23}  \sin^2\Delta $. This gives rise to the 
intrinsic hierarchy  and octant degeneracies:   
\begin{align}
P_{\mu\mu}(\Delta) & = P_{\mu\mu}(- \Delta) \\ 
P_{\mu\mu}(\theta_{23}) & = P_{\mu\mu}(\pi/2 - \theta_{23})\label{8_fold_1}.
\end{align}
This leads to a loss of sensitivity to the hierarchy and octant, 
when the measurement is performed using this channel. 

\item The appearance channel $ P_{\mu e} $ does not have intrinsic 
degeneracies but 
suffers from the combined effect of different parameters, which leads 
to the following set of degeneracies:
\begin{align}
P_{\mu e}(\theta_{13}, \delta_{CP}) & = P_{\mu e}(\theta_{13}^{\prime}, \delta_{CP}^{\prime}
\label{8_fold_2} )
\\
P_{\mu e}(\Delta, \delta_{CP}) & = P_{\mu e}(-\Delta, \delta_{CP}^{\prime}
\label{8_fold_3} ) ~.
\end{align}
Equations.(\ref{8_fold_1}, \ref{8_fold_2}, \ref{8_fold_3}) constitute the 
eight-fold degeneracy  discussed in Ref.\cite{barger}.
\end{itemize}
Recently, it has been discussed that in the context of probabilities 
which are dependent on $\sin^2\theta_{23}$,  the octant degeneracy can 
be generalized to include all values of $\theta_{23}$ in the second octant
\cite{usoctant} and can also be correlated with $\dcp$ 
\cite{usoctant,Minakata:2013eoa}. 
The pattern of parameter degeneracies in the three-dimensional 
$\theta_{23} - \theta_{13}-\dcp$ space arising from the appearance 
probability $P_{\mu e}$ has been discussed in  Ref.\cite{coloma}. This is a continuous 
degeneracy and can be expressed as   
\begin{eqnarray} 
P_{\mu e}(\theta_{23},\theta_{13}, \delta_{CP})  & 
=P_{\mu e}(\theta_{23}^{\prime},\theta_{13}^{\prime}, \delta_{CP}^{\prime}) 
\Rightarrow {\textrm{generalized octant degeneracy.}} 
\end{eqnarray}
However the reactor experiments have measured $\sin^2\theta_{13}$ 
with a high degree of accuracy and future measurements are expected to 
improve it further. This has 
reduced the impact of  
$\theta_{13}$ uncertainty on octant degeneracy to a large extent 
\cite{usoctant}.  
In this paper, we consider another generalized degeneracy which is the
hierarchy-$\theta_{23}$ - $\dcp$ degeneracy:  
\begin{align} 
P_{\mu e}(\theta_{23},\Delta, \delta_{CP})  & =P_{\mu e}(\theta_{23}^{\prime},-\Delta^{\prime}, \delta_{CP}^{\prime}) \Rightarrow 
\mathrm{generalized \hspace{2mm}  hierarchy-\theta_{23}-\dcp \hspace{2mm}}
\mathrm{degeneracy} .
\end{align}
This degeneracy can be observed best  in the test $\theta_{23} - \delta_{CP}$ plane.
Studying it in this fashion allows us to view the degeneracies arising out 
of the remaining unknown parameters in a comprehensive manner. We note that while the hierarchy degeneracy is always discrete, the $\theta_{23}- \dcp$ degeneracy arising out of the appearance channel is continuous. On the other hand, the intrinsic octant degeneracy arising from the $P_{\mu \mu}$ channel is independent of $\dcp$ and discrete in $\theta_{23}$ except for $\theta_{23}$ values close to maximal.   Thus combining 
the survival and conversion probabilities gives rise to disconnected degenerate regions in the $\theta_{23} - \dcp$ plane. We have elaborated on this point in the Appendix. 

In the next subsection, we study the occurrence of the above degeneracies in terms of probabilities for \nova\ and T2K and identify the different possible degenerate 
solutions at the probability level. 

\subsection{Identifying degeneracies at the probability level}

\begin{figure}[H]
\vspace{-1.5cm} 
        \begin{tabular}{lr}
                \hspace*{0.6in}
                \includegraphics[width=0.5\textwidth]{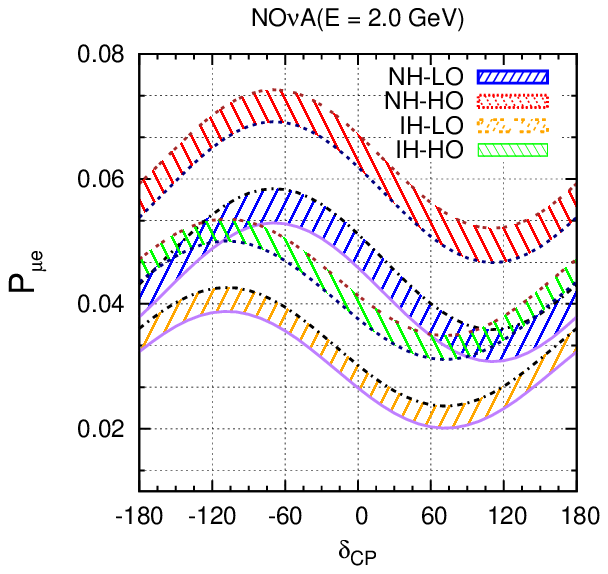}
                &
                \hspace*{-1.0in}
                 \includegraphics[width=0.5\textwidth]{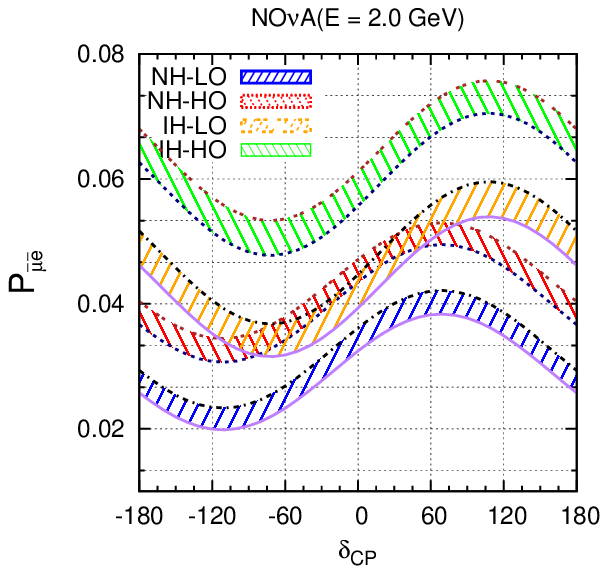} \\
                \hspace*{0.6in}
                \includegraphics[width=0.5\textwidth]{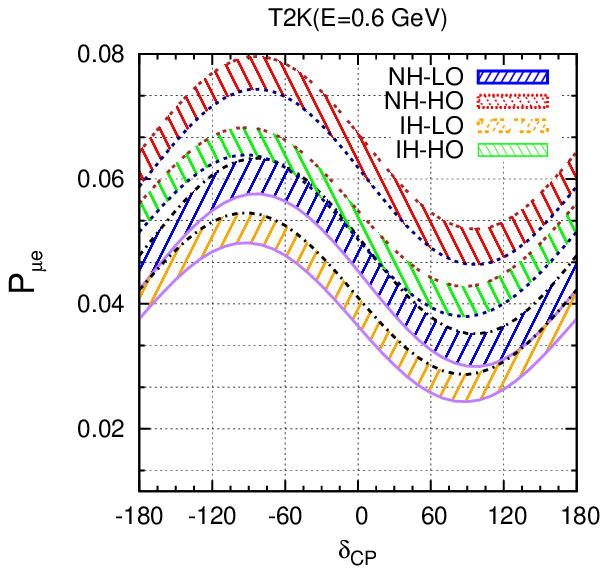}
                &
                \hspace*{-1.0in}
                 \includegraphics[width=0.5\textwidth]{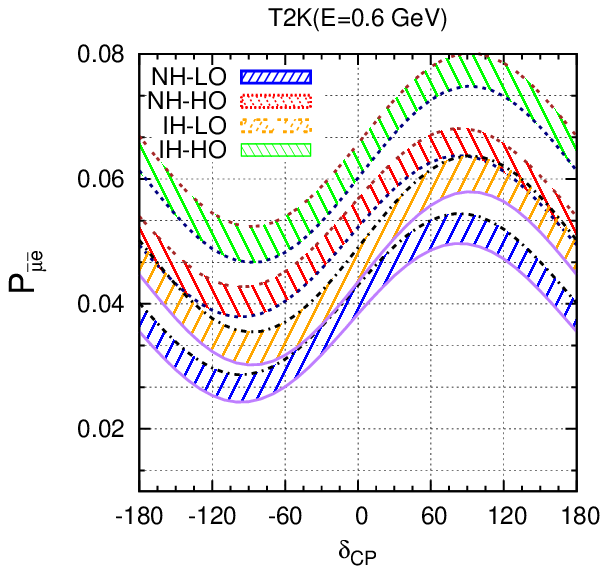}
              
        \end{tabular}
\vspace{0.9cm}
\caption{\footnotesize The oscillation probability  $P_{\mu e}$ 
as a function of $\dcp$. Here, upper row represents oscillation probability for 
\nova\ [L = 810 km], and the lower row represents probability for T2K [L = 295 km]. 
The left panel is for neutrinos, while the right panel is for antineutrinos.
 }
\label{novaprob} 
\end{figure}

Figure~\ref{novaprob} shows the probability $P_{\mu e}$ for neutrinos (left panel) and antineutrinos (right panel) as a function of $\dcp$ for both NH and IH. The plots in the upper panel correspond to \nova\ while those in the lower panel are for T2K. These probabilities are plotted for the energy where  the neutrino flux peaks.
The hatched  area denotes variation over $\theta_{23}$. For the lower (higher) octant, 
we vary $\theta_{23}$ between $39^\circ - 42^\circ$ ($48^\circ - 51^\circ$). 
This is a good assumption for $\theta_{23}$ not too close to its maximal
value, for the purpose of illustrating the physics,  since for a given $\theta_{23}$(true), the disappearance channel anyway excludes values away from $\theta_{23}(\mathrm{true})$ and $\pi/2 - \theta_{23}(\mathrm{true})$. 
Thus, these plots implicitly assume information from the disappearance
channel. From Fig.~\ref{novaprob}, the following points can be noted.\\ 
\vspace{1cm} 
$For neutrinos -$ \\ 
(i) The NH probabilities are higher than the IH probabilities. This is 
because of enhanced matter effect for neutrinos for NH in the Earth's matter.\\   
(ii) For both NH and IH the probabilities are higher in the LHP. \\
(iii) The probabilities for higher octant are  higher for both NH and IH. \\
$While, for ~ antineutrinos -$ \\
i) \enquote{A} changes its sign, and IH probabilities become higher than NH.\\
ii) The flip in sign of $\dcp$ makes both the NH and IH probabilities 
higher in the UHP.\\
iii) Like neutrinos, the probabilities for a higher
octant remains higher for both NH and IH. \\

For both neutrinos and antineutrinos the lowest line in the LO (HO)  
band corresponds to $39^\circ$(48$^\circ$), while the highest point corresponds to $42^\circ$ ($51^\circ$),  due to the $\sin^2 \theta_{23}$ dependence of the leading order term.

The overlapping regions between various curves at a specific value of the
$\dcp$ indicate the degeneracy  
occurring {\it for the right value of $\dcp$},  
while the same value of probability for  different $\dcp$ values  
denotes degeneracy occurring {\it at wrong values of $\dcp$}. 
Clearly, the former would correspond to solutions with the wrong hierarchy and/or
octant with right $\dcp$, 
while the latter would correspond to the solutions
with the wrong hierarchy and/or octant and  wrong $\dcp$. 
The wrong-CP degenerate solutions corresponding to a given true value of $\theta_{23}$ 
and $\dcp$ can be obtained from the 
probability figures by drawing a horizontal line through this point. 
The different intersection points of this line with the probability bands
are degenerate as they share the same value of probability. 
However, the degenerate solutions occurring for 
a particular $\theta_{23}$ (true) which is not in the vicinity
of  $\pi/2 -\theta_{23}$ (true) in the opposite octant  will be excluded 
by the disappearance channel, and the
occurrence of these solutions in the test $\theta_{23}-\dcp$ plane will 
depend on the $\theta_{23}$ precision of the disappearance channel.

Below, we explain the occurrence of the different degenerate combinations 
of \{hierarchy, $\theta_{23}$, $\dcp$\} taking 
the \nova\ neutrino probabilities (the top-left panel) as reference unless 
otherwise mentioned: 
\begin{enumerate}
 \item The overlapping regions between the NH-LO (blue) and IH-HO (green) 
 bands around 
 $\dcp= -120^\circ$ and $90^\circ$ give rise to WH-WO-R$\dcp$ degenerate 
 solutions in the probability.  
However, for antineutrinos these bands are well separated. 
Thus, combining \nova\ neutrino and antineutrino data can help in 
removing these solutions. 
 
 \item The probability corresponding to UHP of the NH-LO (blue) band can 
 be the same as those for the LHP of the IH-LO (yellow) band for $\theta_{23} = 39^\circ$. 
This can give rise to WH-RO-W$\dcp$ solutions. Note that 
this degeneracy is present in the antineutrinos for the same values of 
$\dcp$. Thus, for true NH, UHP (i.e. $0^\circ < \dcp < 180^\circ$) is the
unfavorable half-plane of $\dcp$, and this degeneracy cannot be resolved 
by using \nova\ data alone. For T2K, the probability exhibits a sharper variation with $\dcp$, and hence this degeneracy is less pronounced between the UHP and LHP. 
Hence, the addition of T2K data to \nova\ can be helpful in
removing this degeneracy. For LHP (i.e. $-180^\circ < \dcp < 0^\circ$),
which is the favorable half-plane 
of $\dcp$ in NH, there is no  WH-RO-W$\dcp$ solution for both \nova\ and T2K.  
 
 \item  For $\dcp \in$ UHP, the NH-HO (red) band can share same value 
 of probability  with NH-LO (blue) band for $\dcp \in$ LHP. This is the reason for the 
 RH-WO-W$\dcp$ solution. For antineutrinos, the degeneracy is seen to be between NH-HO-LHP and NH-LO-UHP.  Thus, a combination of neutrinos and antineutrinos helps 
to remove this degeneracy. 
 
  \item The WH-WO-W$\dcp$ solution can be observed along the isoprobability 
  line that   intersects the NH-LO (blue) and IH-HO (green) bands at different values 
  of $\dcp$.   This degeneracy can be seen for instance between \{NH, $39^\circ$, $-180^\circ$\}   and \{IH, $51^\circ$, $0^\circ$\}.   Again, the antineutrino probability does not suffer from this degeneracy, and thus combining neutrino and antineutrino data can be helpful in removing these solutions. 
  
\item  One can also have RH-RO-W$\dcp$ solutions as a result of a so called \enquote{intrinsic CP degeneracy} that occurs for the same hierarchy and same value of $\theta_{23}$ but at a different value of $\dcp$. This is due to the harmonic 
dependence of the probability on $\dcp$. For instance, within the NH-HO (blue) band, 
$\dcp=0^\circ$ and $\dcp \approx -135^\circ$ have the same value of probability for
$\theta_{23} = 39^\circ$.  However, for antineutrinos, this occurs for $\dcp=0^\circ$ and $\dcp = +135^\circ$. Thus, a combination of neutrino and antineutrino data can help to get rid of this degeneracy. This can also be seen to occur for T2K, for \{NH, $48^\circ$,  $-180^\circ$\} and \{NH, $48^\circ$,  $0^\circ$\}. 
For T2K, since the flux peak coincides with the probability peak, 
the CP-dependent term is proportional to $\sin\dcp$ and thus 
this degeneracy occurs for $\dcp$ and $\pi-\dcp$ \cite{Minakata:2013eoa}. 
For \nova, since the flux and the probability peak are not at the same 
energy, the degeneracy does not correspond exactly to $\dcp$ and $\pi-\dcp$.
It is interesting to note that this degeneracy does not occur for 
$\dcp=\pm 90^\circ$.
  
\end{enumerate}

Thus, among the eight solutions listed above, only the WH-RO-R$\dcp$ 
and RH-WO-R$\dcp$ solutions do not exist even at the probability level. 
The above description is in terms of probabilities without including 
any experimental errors. At the event level, many of these may not 
appear as discrete degeneracies. In particular, for a C.L. beyond the reach of a particular experiment's precision, the different discrete degenerate solutions merge, 
and the degeneracy becomes continuous.

 Note that another way to understand the degeneracies is the 
biprobability ellipses in the $P_{\mu e} - P_{\bar{\mu} \bar{e}}$  plane \cite{Minakata:2001qm}.
This requires a single plot for neutrinos and antrineutrinos. 
However, each point on these ellipses corresponds to different values of 
$\dcp$ which cannot be read off from the plots. The probability band plots 
presented in this paper provide a complementary way to visualize the occurrence of 
the degeneracies at different CP values, and one can readily identify 
the wrong and right $\dcp$ solutions which are the main 
focus of this work.

\subsection{Identifying degeneracies at the event level}

\begin{figure}[H]
        \begin{tabular}{lr}
                \includegraphics[width=0.45\textwidth]{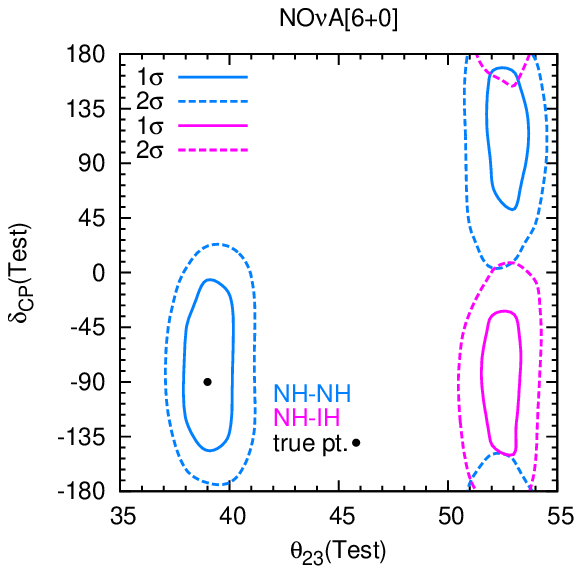}
                 
                \hspace*{-1.0in}
                 \includegraphics[width=0.45\textwidth]{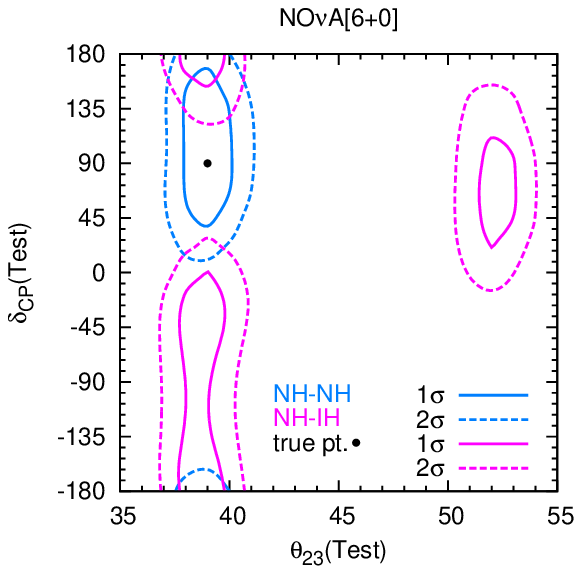}
                 \hspace*{-1.0in}
                 \includegraphics[width=0.45\textwidth]{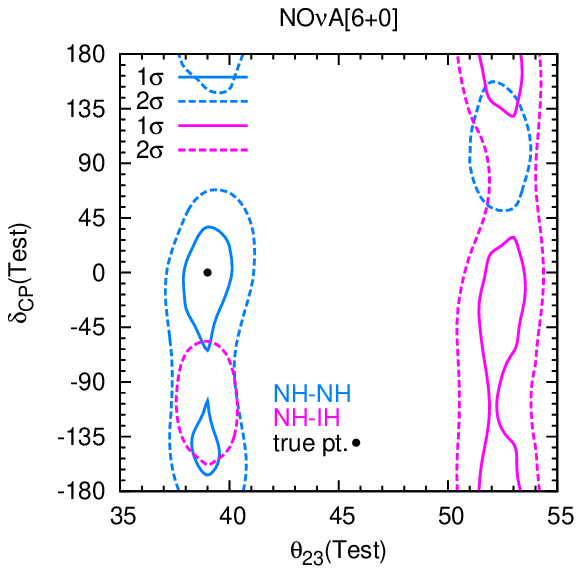}\\

                \includegraphics[width=0.45\textwidth]{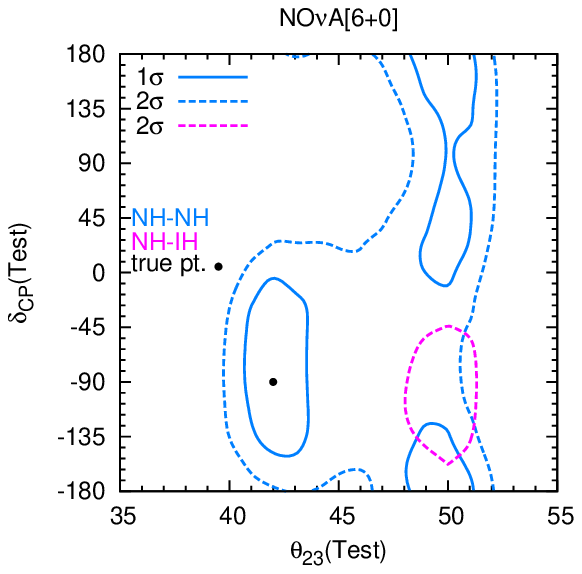}
                 
                \hspace*{-1.0in}
                 \includegraphics[width=0.45\textwidth]{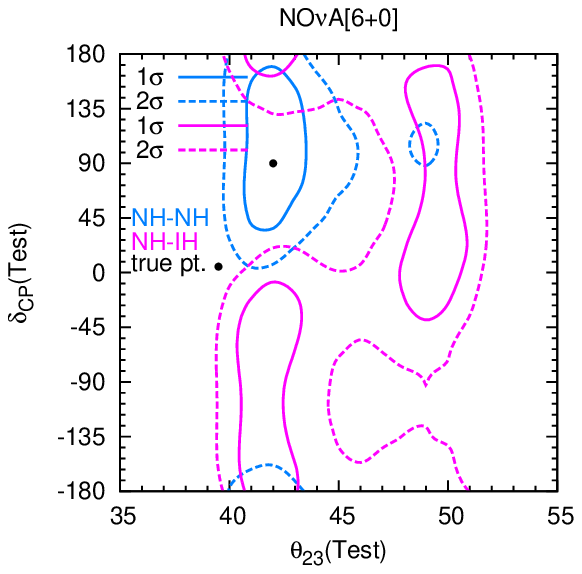}
                 \hspace*{-1.0in}
                 \includegraphics[width=0.45\textwidth]{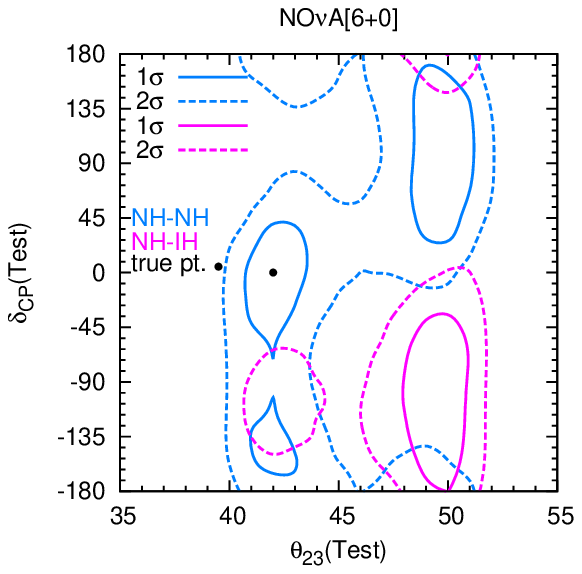}\\

                \includegraphics[width=0.45\textwidth]{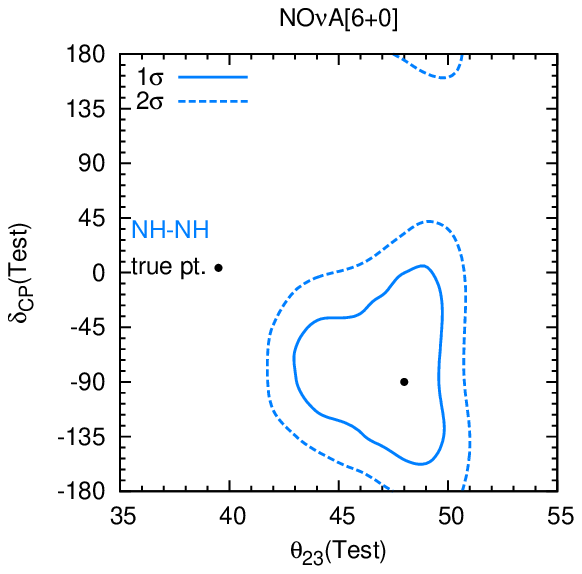}

                \hspace*{-1.0in}
                 \includegraphics[width=0.45\textwidth]{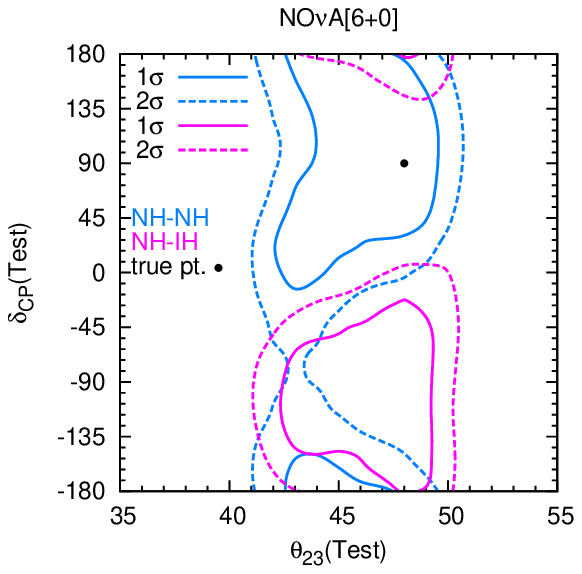}
                 \hspace*{-1.0in}
                 \includegraphics[width=0.45\textwidth]{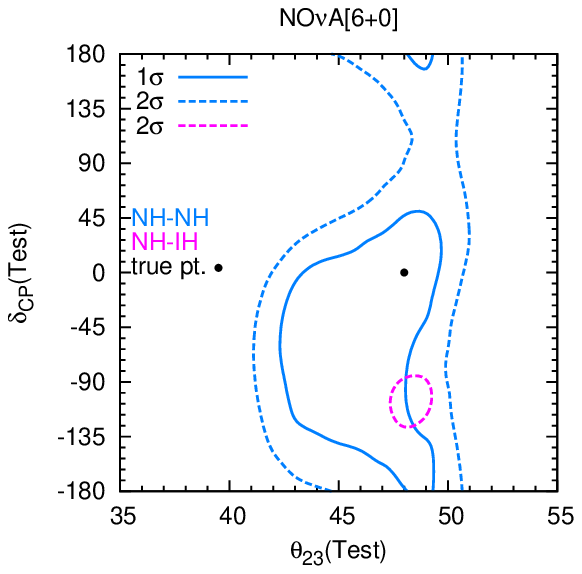}
\\
                 
                \includegraphics[width=0.45\textwidth]{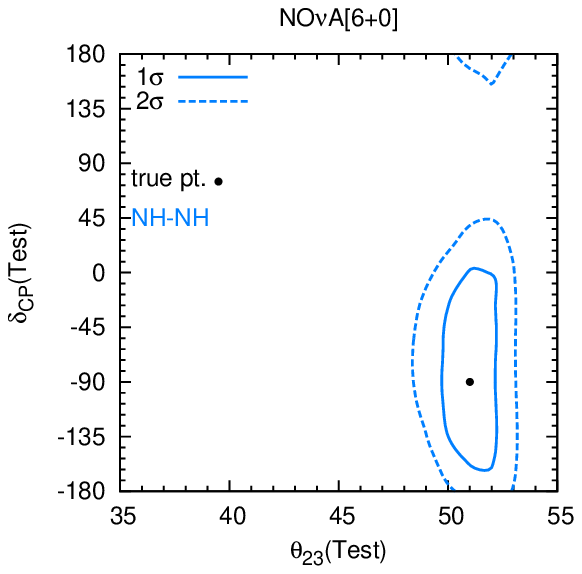}
                
                \hspace*{-1.0in}
                 \includegraphics[width=0.45\textwidth]{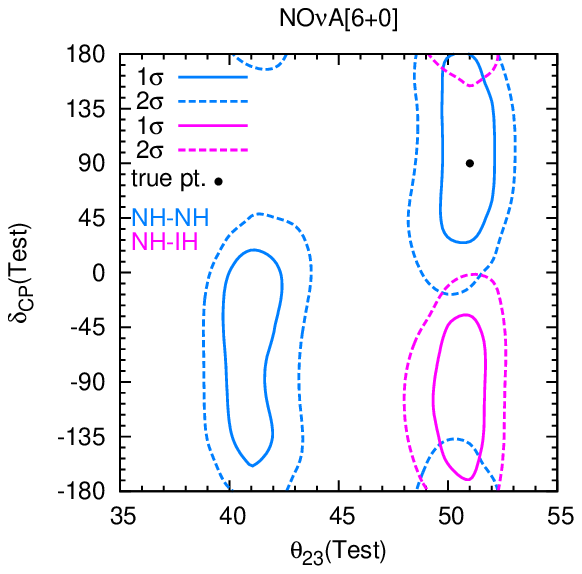}
                 \hspace*{-1.0in}
                 \includegraphics[width=0.45\textwidth]{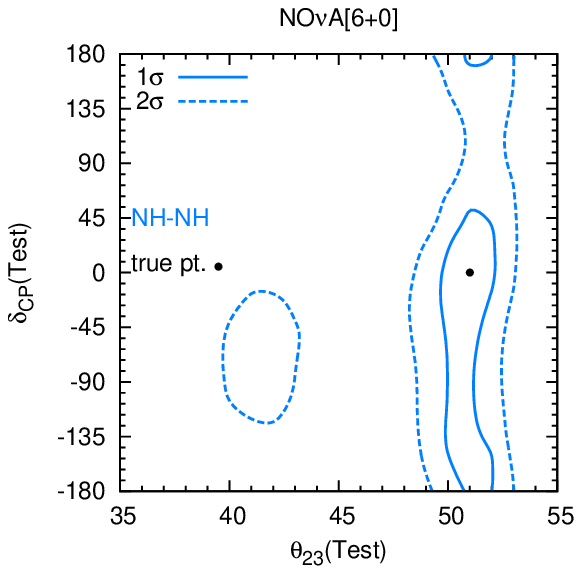}

                      \end{tabular}
\vspace{1.0cm}
\caption{\footnotesize Contour plots for  \nova[6+0] with true values of 
$ \theta_{23} = 39^\circ, 42^\circ, 48^\circ, 51^\circ $ in successive rows.  
The three  
columns
correspond to $\delta_{CP} = -90^\circ,90^\circ,0^\circ $ respectively
.}
\label{only_nu}
\end{figure}

To show the occurrence of the different degeneracies at the event level, 
in Fig.~\ref{only_nu} 
we present a set of contour plots in the $\theta_{23} - \dcp$ 
test-parameter plane assuming only neutrino run (6 years) of \nova, which is denoted as 
[6+0]. We note that the proposed run time of \nova\ is 3+3. 
However in this section we intend to identify  which of 
the different degenerate solutions 
discussed in Table \ref{degneracy_table} can arise in the $\theta_{23}-\dcp$ plane. 
Since the wrong-octant solutions disappear including the antineutrino
run,  the 6+0 case is the best option for visualizing all the possible
degenerate solutions. The 3+3 case is discussed in the next section.
These sets of plots also show the role of statistics which can give enhanced 
precision  $vis$-{\`a}-$vis$ the antineutrino
run which helps in resolving degeneracies.
Similar plots can also be drawn for the T2K 8+0 case.
However, for T2K,  since the hierarchy sensitivity is much less, 
the possibility of getting  continuous regions instead of discrete 
degenerate solutions is more. Thus, the  different degenerate 
solutions cannot be visualized so distinctly, and in this section, our main aim is to 
identify the different degenerate solutions in the $\theta_{23}-\dcp$ plane. This 
can be done better with \nova\ 6+0 as the illustrative example. 
These plots are drawn assuming true  hierarchy to be NH and different choices 
of true  values of $\theta_{23}$ and $\dcp$. In this and all the other subsequent 
figures the successive rows are for $\theta_{23} = 39^{\circ}, 42^\circ, 48^\circ, 51^{\circ}$. The true $\dcp$ values chosen are $\pm 90^\circ$
corresponding to maximum CP violation and $0^\circ$ corresponding to CP conservation.
The blue contours correspond to the right hierarchy and magenta curves
correspond to  the wrong-hierarchy. 

The first column of Fig.~\ref{only_nu} shows the degenerate solutions for 
$\dcp = -90^\circ$ for \nova\ running only in  the neutrino mode. 
For $\theta_{23} = 39^\circ$, apart from the true solution, 
RH-WO-W$\dcp$ and WH-WO-R$\dcp$ solutions are observed  in the upper and lower 
right quadrants respectively. The RH-WO-W$\dcp$ solution is also seen for $\theta_{23}= 42^\circ$. For this case, at $\dcp=-90^\circ$, 
the uppermost  point of the blue band in the \nova\ neutrino probability, in 
Fig.~\ref{novaprob} one can see that the same value of probability is possible for NH-HO (red band) near $\dcp = +45^\circ$ and $\pm 180^\circ$. This explains the  shape of the 
allowed zone -- wider at these values and narrower at $90^\circ$. 
The WH-WO-R$\dcp$ solution is seen only at a 2$\sigma$ level for $\theta_{23}=42^\circ$.
This can be understood by observing that  
the points $42^\circ$ (the upper tip of the blue band) 
and $48^\circ$ (the lower tip of the green band) are more separated 
as compared to $39^\circ$ (the lower tip of the blue band) and 
$51^\circ$ (the upper tip of the green band). 
For $\theta_{23}$ in the higher octant ($48^\circ$ and $51^\circ$)
there are no spurious wrong-hierarchy solutions  even with only neutrinos. 
This is because for NH, $\theta_{23}$ in the higher octant 
and  $-90^\circ$ correspond to the maximum probability 
for neutrinos and this cannot be matched by any other 
combination of parameters. Hence, no degenerate solutions appear and 
only the neutrino run for \nova\ suffices to give an allowed area only near the 
true point. Note that the contours  for $48^\circ$  extend to the 
wrong octant also. However, (here and elsewhere) 
this is not due to any degenerate behavior of 
the $P_{\mu e}$ probability but due to the poor $\theta_{23}$ 
precision of the $P_{\mu \mu}$ channel  near maximal mixing. 

The second column represents $\dcp=+90^\circ$. In this case,  we observe 
a WH-WO-R$\dcp$ solution for  both $\theta_{23}=39^\circ$ and $42^\circ$.
This can be understood from the intersection of the blue and green bands 
in  Fig.~\ref{novaprob} close to $\dcp = 90^\circ$ in the UHP. We also get a  WH-RO-W$\dcp$ region in the LHP. For $42^\circ$, since the $\theta_{23}$ precision coming 
from the disappearance channel is worse, at $2\sigma$ both these solutions merge, and a discrete degenerate region is not obtained. For $ \theta_{23} = 51^\circ$ from Fig.~\ref{novaprob}, 
we see that the point \{NH, $+90^\circ$, $51^\circ$\} in the red band intersects the blue band around \{NH, $-90^\circ$, $39^\circ$\} giving a  RH-WO-W$\dcp$ solution. 

\hspace{2cm} 
 A WH-RO-W$\dcp$ solution is also obtained in this case in the LHP.   
Similar regions are also obtained for $\theta_{23}=48^\circ$. 
However, the RH-WO-W$\dcp$ solution merges with the true solution at $2\sigma$ level.

For  $\dcp=0^\circ$,  a discrete RH-RO-W$\dcp$ solution is seen to be allowed at 1$\sigma$ 
for $\theta_{23} \in $ LO. This is  due to the intrinsic CP degeneracy as discussed in the context of probabilities. But at 2$\sigma$, due to the poor 
$\dcp$ precision this degeneracy becomes continuous and the whole LHP becomes allowed. For $\theta_{23}$ belonging to the higher octant  larger statistical errors
are involved  as compared to $\theta_{23} \in$ LO,  and this degeneracy
appears as continuous in the LHP even at 1$\sigma$, and at 2$\sigma$ the full 
$\dcp$ range becomes allowed. For $\theta_{23}=39^\circ$ and $42^\circ$, we also see wrong-hierarchy solutions appearing in the wrong octant. From the probability figure, we identify that this degeneracy occurs around $\dcp= -30^\circ, -180^\circ, 180^\circ$ 
for $\theta_{23} = 39^\circ$ which allows the LHP of $\dcp$ at 1$\sigma$ and 
the whole $\dcp$ range at 2$\sigma$. For $\theta_{23} = 42^\circ$, this
degeneracy is seen to  occur around $\dcp = -90^\circ$ giving distinct 
degenerate solutions at the $1\sigma$ and $2\sigma$ levels. 
For $\theta_{23} = 42^\circ$, a discrete RH-WO-W$\dcp$ solution  appears at $1\sigma$.  
From Fig.~\ref{novaprob}, it is seen that 
\{NH, $42^\circ$, $0^\circ$\} has the same value of probability  
corresponding to \{NH, $48^\circ$,   $90^\circ$\}. 
At $2\sigma$, this merges with the  RH-RO-W$\dcp$ solution.
For $\theta_{23} = 39^\circ$, this solution appears as a 2$\sigma$ 
allowed patch around \{NH, $51^\circ$,  $90^\circ$\}.  
From Fig.~\ref{novaprob}, it can be seen that
the above points are not exactly degenerate in terms of 
probability but due to lack of precision, they become allowed.  
For a similar reason, the $2\sigma$ patch with wrong-hierarchy  
appears in the right octant for $\theta_{23}= 39^\circ$ and $42^\circ$. 
For $\theta_{23} = 51^\circ$ a right-hierarchy patch occurs with wrong octant. 
For $\theta_{23} = 48^\circ$, because of proximity to maximal mixing, 
the true parameter space also extends to the wrong-octant. 
In general, we see that the CP precision is poorer for $\dcp=0^\circ$ at this stage.
This is due to the unresolved degeneracies for $\dcp=0^\circ$ which 
lead to multiple allowed regions and continuous bands at $2\sigma$.
  
\subsection{Successive resolution of degeneracies with data from different 
experiments}

In this section, first we show the status of the above degenerate regions
when \nova\ runs in [3+3] configuration.
We then study the combined potential of \nova[3+3] and T2K[8+0] 
in resolving the degeneracies. Finally, we add the atmospheric neutrino data 
from ICAL and show its impact. 

 It is well known that due to Earth matter effects and the presence 
of an antineutrino component in the atmospheric neutrino flux, ICAL can play 
a prominent role in resolving the      
hierarchy and octant degeneracies \cite{ino2d, ino3d}. 
Since there is no $\dcp$ 
dependence in ICAL, the  hierarchy and octant sensitivities
are independent of $\dcp$. 
This $\dcp$-independent $\chi^2$ 
adds to the  $\chi^2$ for \nova\ and T2K in the degenerate region, 
and the wrong-hierarchy and wrong-octant solutions can be resolved. 
This aids in improving both the octant and $\dcp$ sensitivities of T2K and \nova\ 
\cite{usoctant,mono_ical, ourlongcp}.
Note that this is a synergistic effect and the combined sensitivity is 
better than that obtained by adding individual $\chi^2$ values.

The advantage offered by the atmospheric neutrino detector ICAL and
the synergy between the 
various experiments in removing the degeneracies  
can be seen from these plots. 
It is noteworthy that the allowed area in the test
$\theta_{23} - \dcp$ plane also gives an idea about the 
precision of these two parameters.  

\begin{center}
 \begin{figure}[H]
        \begin{tabular}{lr}
\vspace{-5mm}        
                \includegraphics[width=0.45\textwidth]{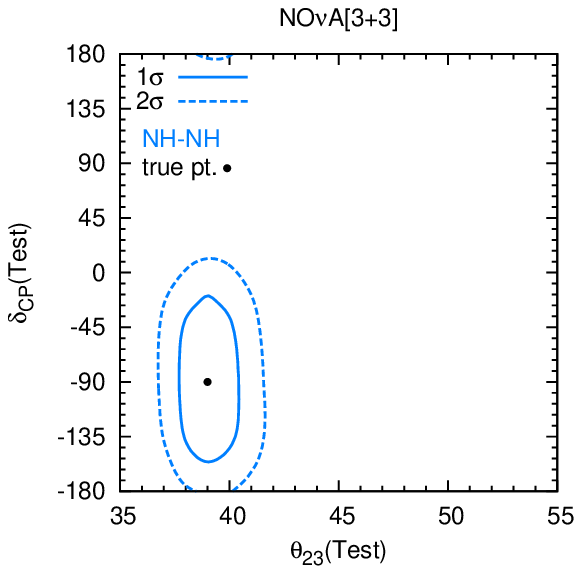}                 
                \hspace*{-1.0in}
                 \includegraphics[width=0.45\textwidth]{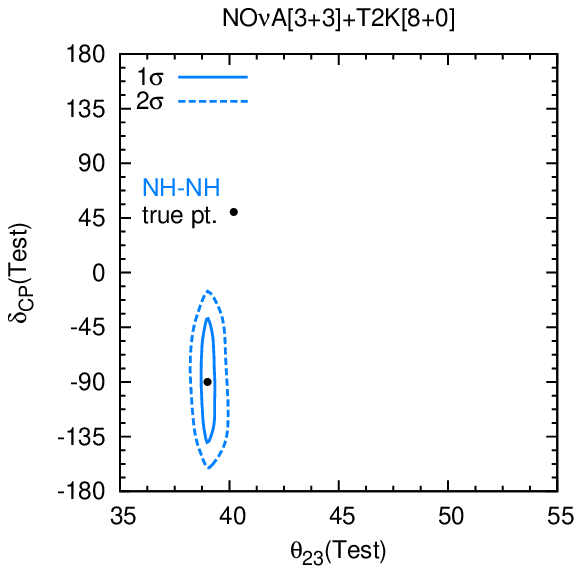}
                 \hspace*{-1.0in}
                 \includegraphics[width=0.45\textwidth]{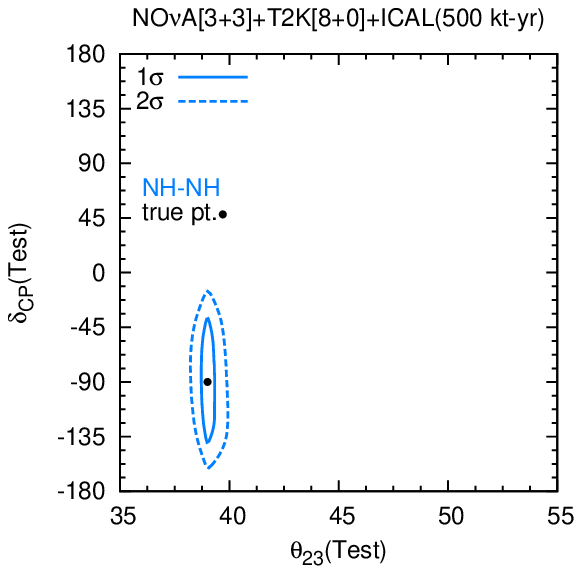}\\
\vspace{-5mm}
                \includegraphics[width=0.45\textwidth]{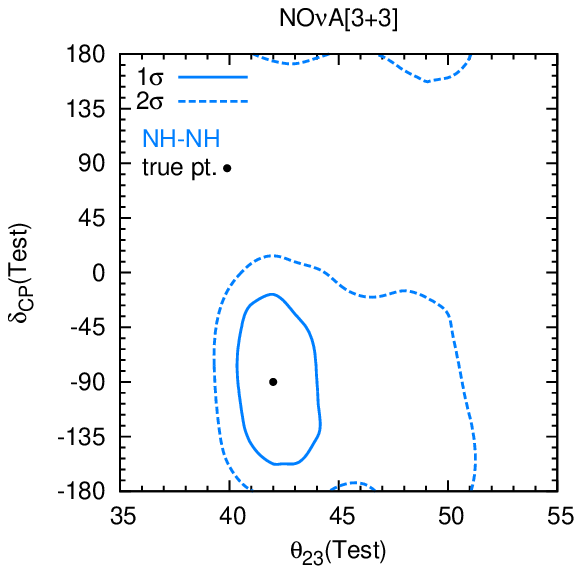}
                 
                \hspace*{-1.0in}
                 \includegraphics[width=0.45\textwidth]{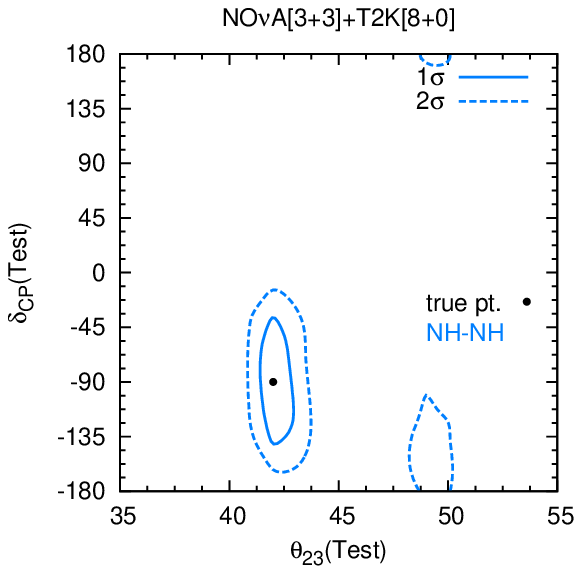}
                 \hspace*{-1.0in}
                 \includegraphics[width=0.45\textwidth]{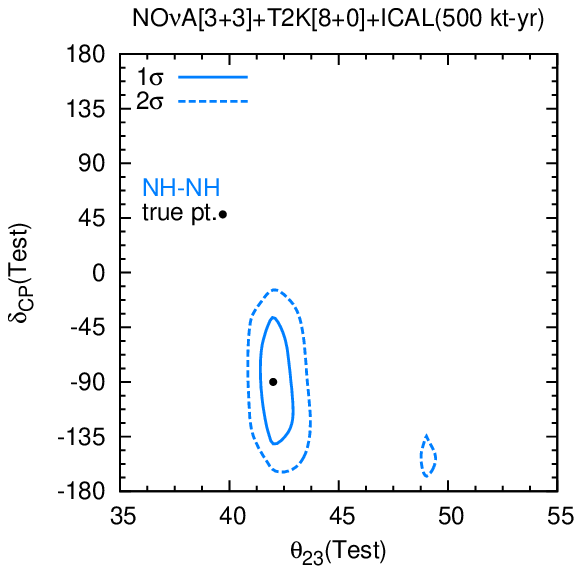}\\
\vspace{-5mm}                  

                \includegraphics[width=0.45\textwidth]{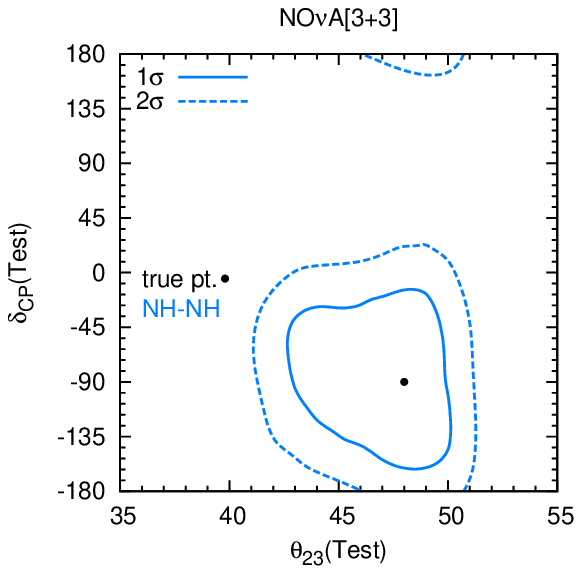}

                \hspace*{-1.0in}
                 \includegraphics[width=0.45\textwidth]{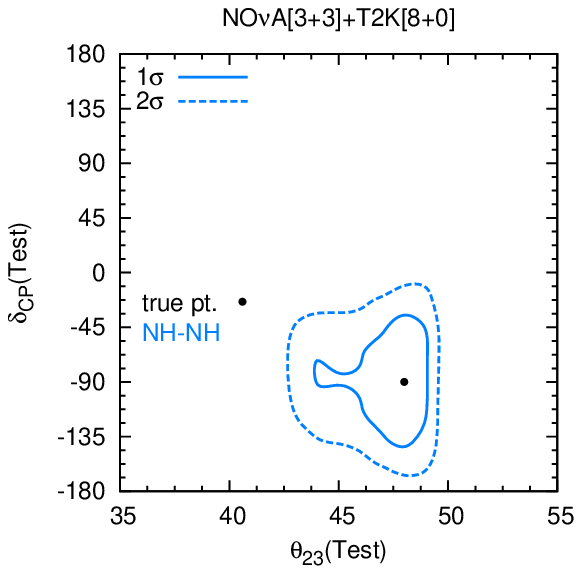}
                 \hspace*{-1.0in}
                 \includegraphics[width=0.45\textwidth]{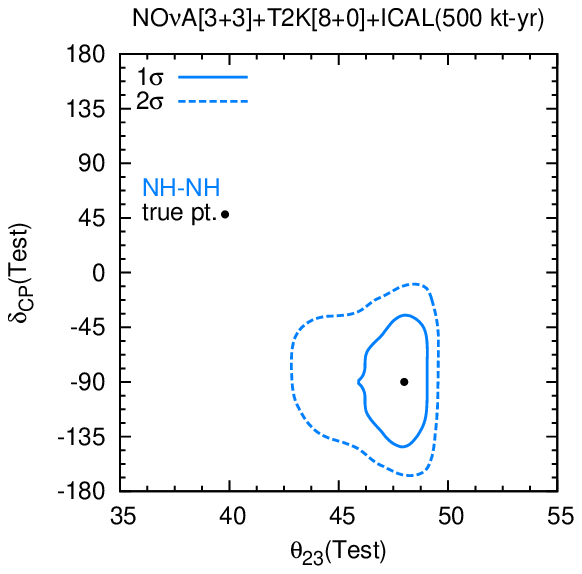}
\\ 
\vspace{-5mm}                 
                \includegraphics[width=0.45\textwidth]{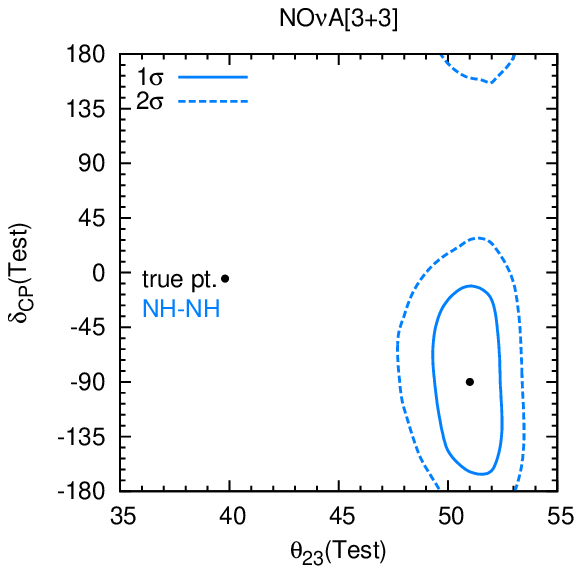}
                
                \hspace*{-1.0in}
                 \includegraphics[width=0.45\textwidth]{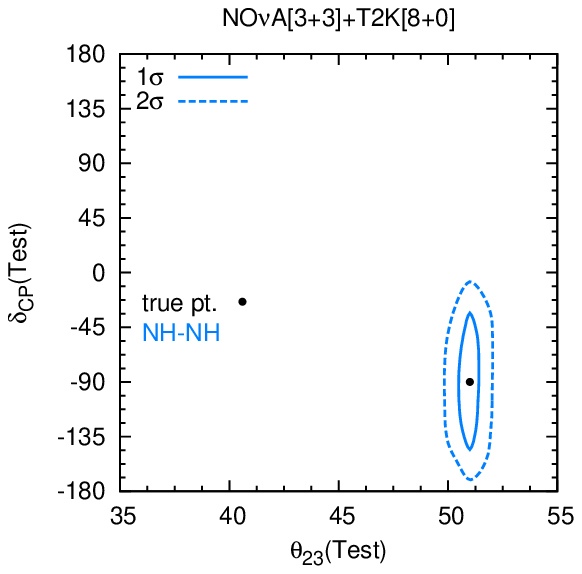}
                 \hspace*{-1.0in}
                 \includegraphics[width=0.45\textwidth]{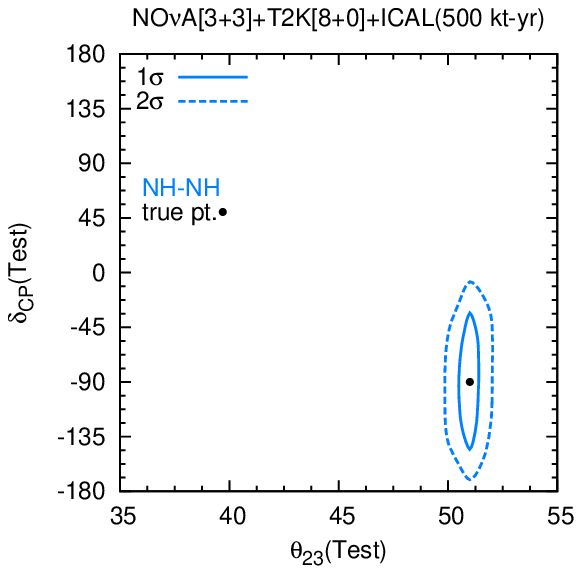}

                      \end{tabular}
\vspace{1.3cm}
\caption{\footnotesize Contour plots in the test $\theta_{23} - \dcp$ plane for true $\dcp= -90^\circ$ and true $\theta_{23}=39^\circ, 42^\circ, 48^\circ$ and
 $51^\circ$ in successive rows. The first  column is for \nova[3+3]. The second and third columns are for \nova[3+3] + T2K [8+0] and \nova[3+3]+T2K[8+0]+ICAL respectively.
}
\label{fig:cp_-90}
\end{figure}
 \end{center}

\begin{figure}[H]
        \begin{tabular}{lr}
 \vspace{-5mm}       
                 \includegraphics[width=0.44\textwidth]{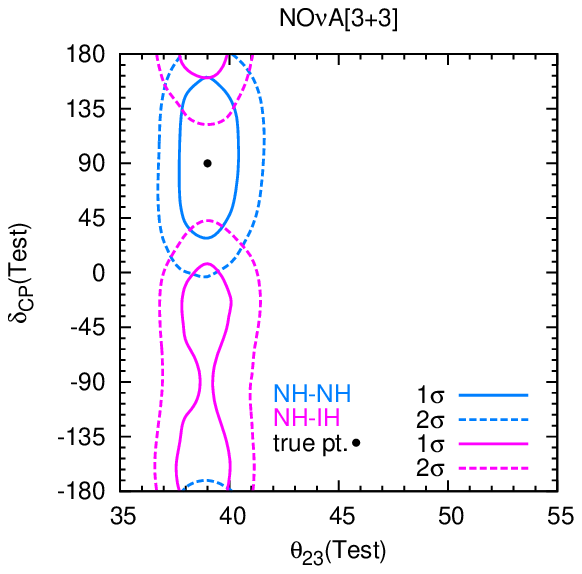}
                 \hspace*{-1.0in}
                \includegraphics[width=0.44\textwidth]{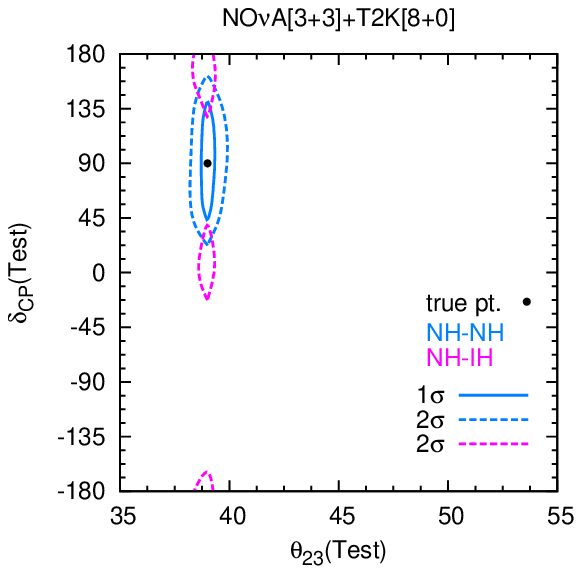}
                 \hspace*{-1.0in}
                 \includegraphics[width=0.44\textwidth]{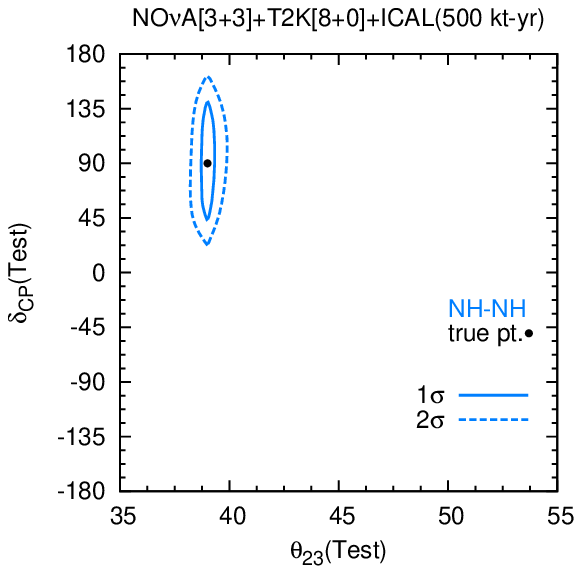}\\
\vspace{-5mm}                 
                 %
                 %
                 \includegraphics[width=0.44\textwidth]{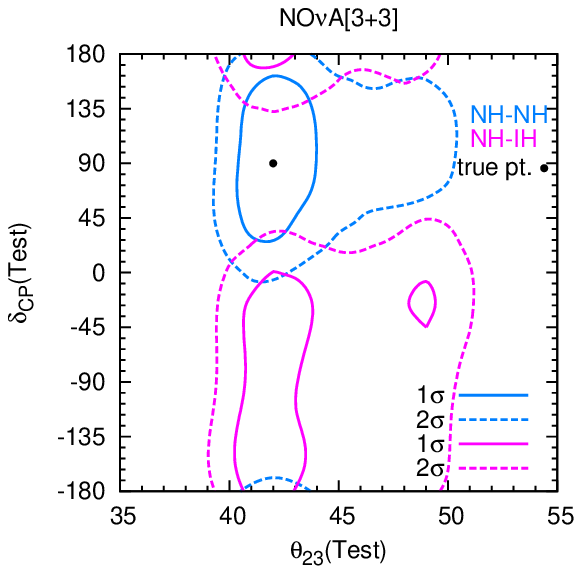}
                 \hspace*{-1.0in}
                \includegraphics[width=0.44\textwidth]{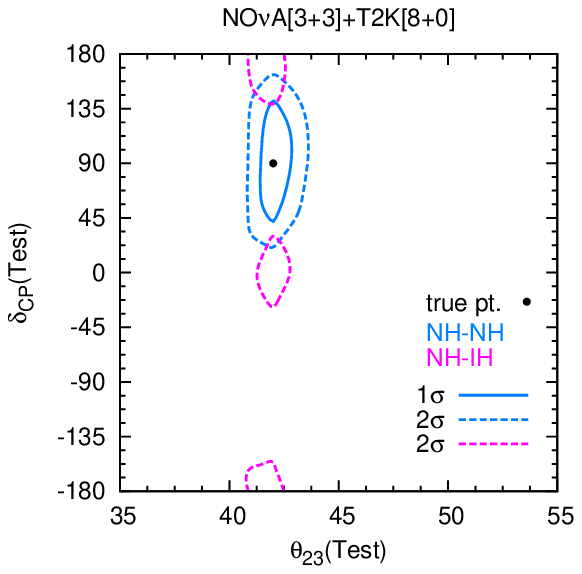}
                 \hspace*{-1.0in}
                 \includegraphics[width=0.44\textwidth]{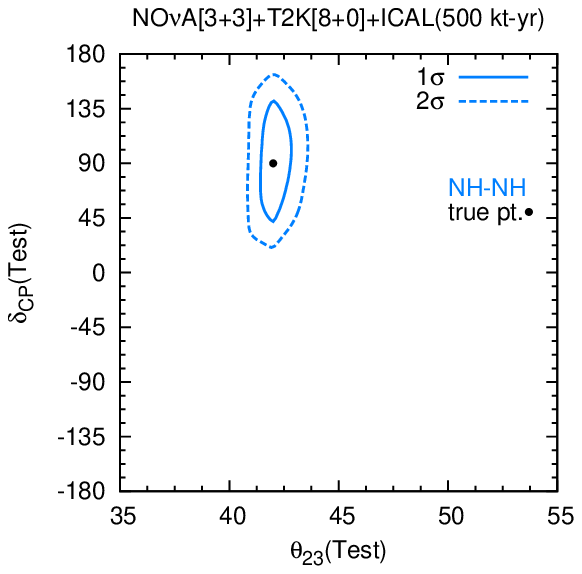} \\
 \vspace{-5mm}                
                 %
                 %
                 \includegraphics[width=0.44\textwidth]{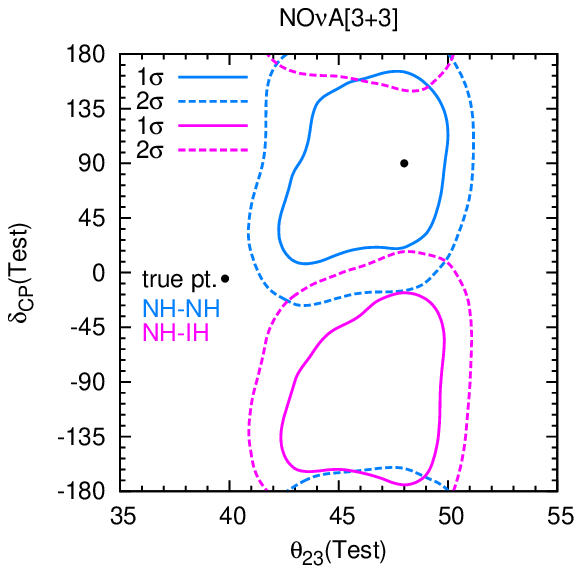}
                 \hspace*{-1.0in}
                \includegraphics[width=0.44\textwidth]{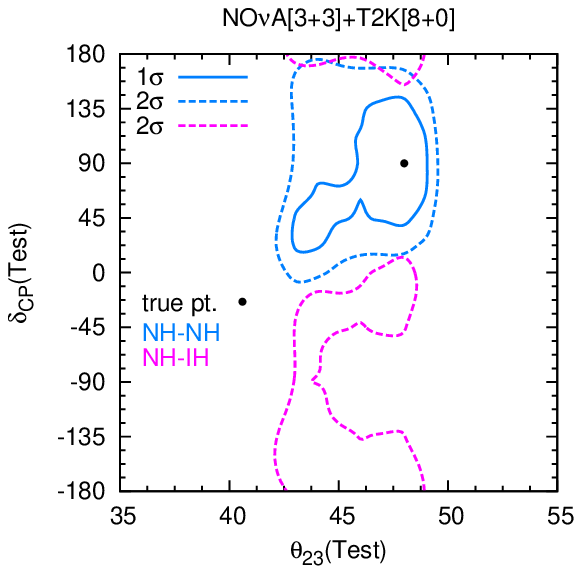}
                 \hspace*{-1.0in}
                 \includegraphics[width=0.44\textwidth]{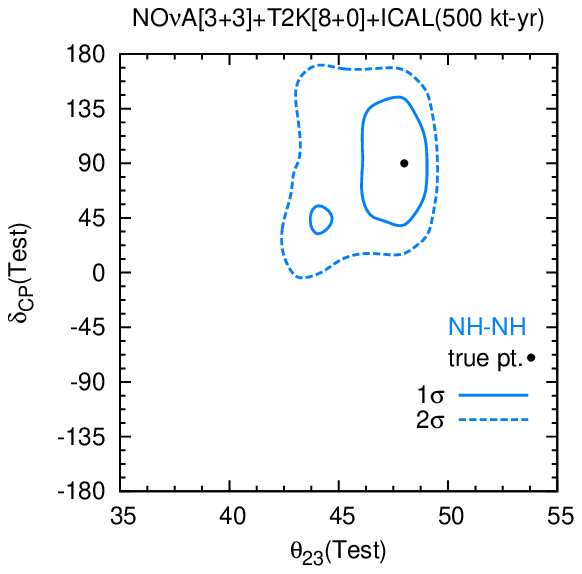}
                 \\
\vspace{-5mm}                 
                 %
                 %
                 \includegraphics[width=0.44\textwidth]{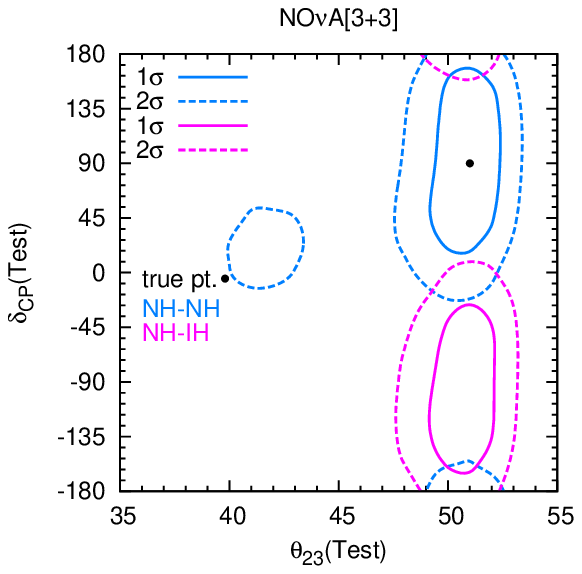}
                 \hspace*{-1.0in}
                \includegraphics[width=0.44\textwidth]{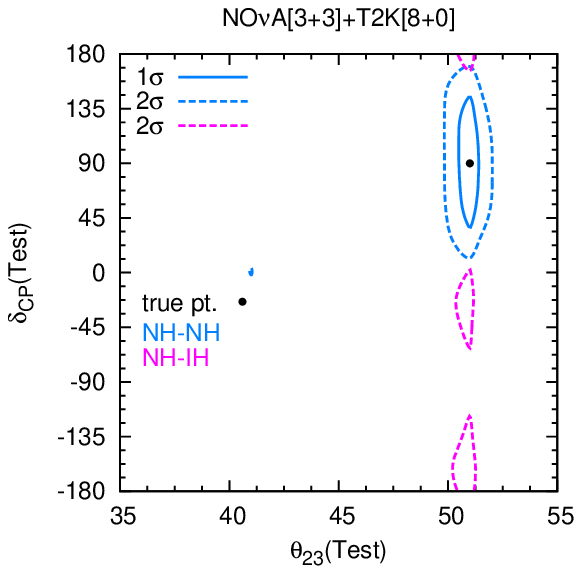}
                 \hspace*{-1.0in}
                 \includegraphics[width=0.44\textwidth]{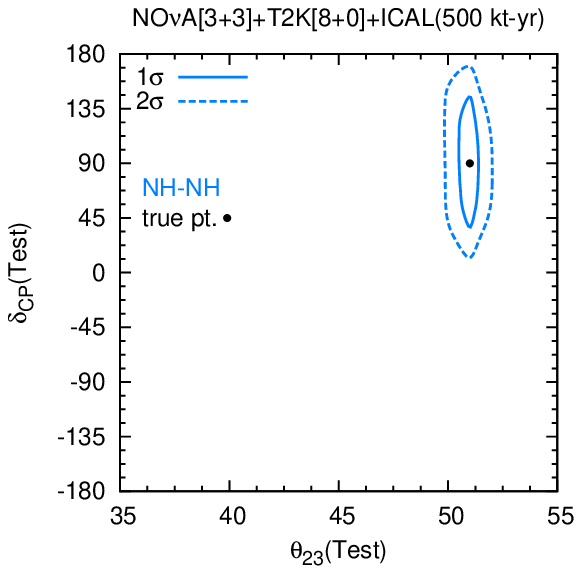}
                 
        \end{tabular}
 \vspace{1.3cm}
 \caption{\footnotesize Same as in Fig.\ref{fig:cp_-90} but for true $\dcp= +90^\circ$ .}
\label{fig:cp_90} 
\end{figure}

\begin{figure}[H]
        \begin{tabular}{lr}
\vspace{-5mm}        
                 \includegraphics[width=0.44\textwidth]{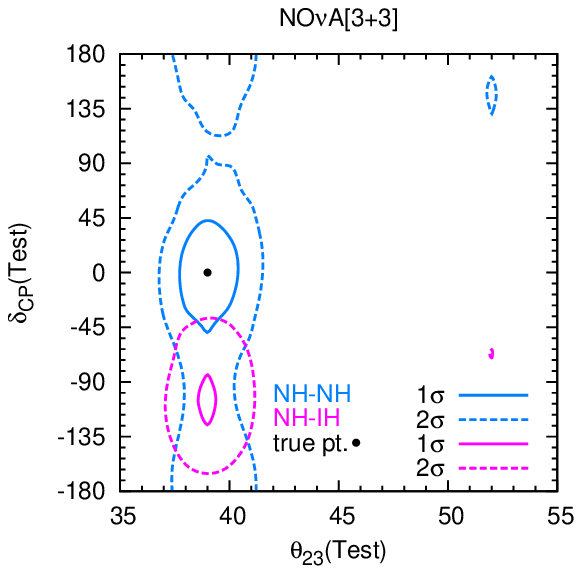}
                 \hspace*{-1.0in}
                \includegraphics[width=0.44\textwidth]{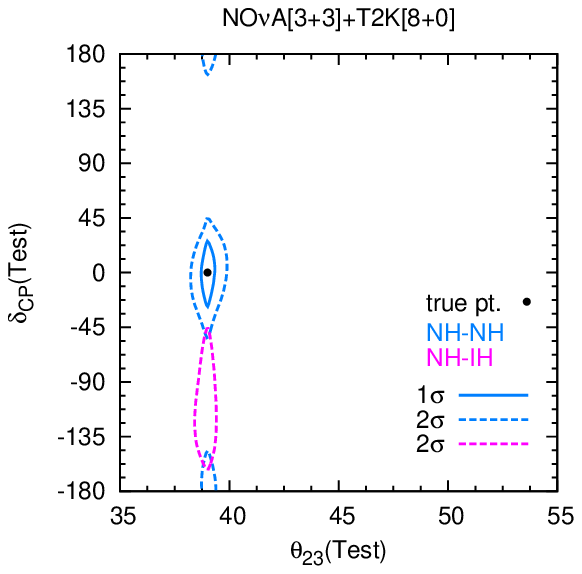}
                 \hspace*{-1.0in}
                 \includegraphics[width=0.44\textwidth]{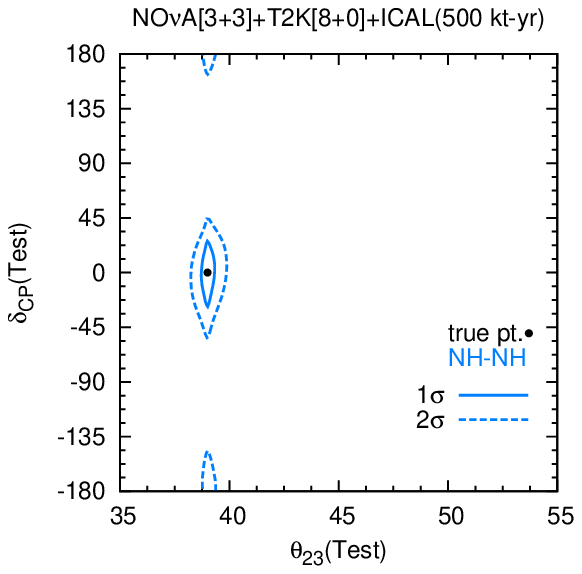}\\
\vspace{-5mm}                 
                 %
                 %
                 \includegraphics[width=0.44\textwidth]{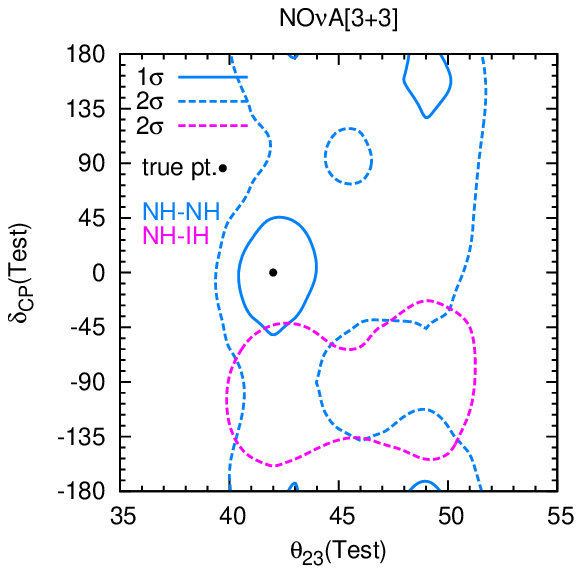}
                 \hspace*{-1.0in}
                \includegraphics[width=0.44\textwidth]{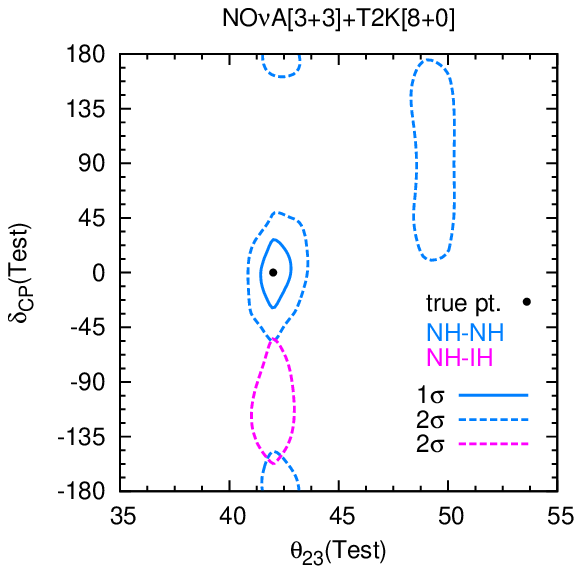}
                 \hspace*{-1.0in}
                 \includegraphics[width=0.44\textwidth]{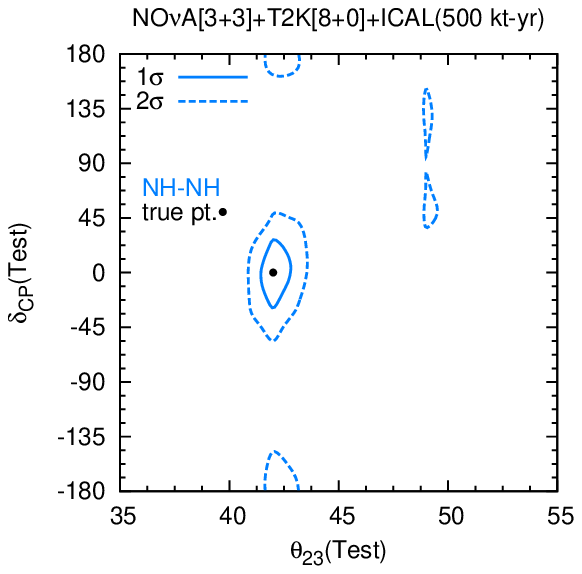}
                 \\
\vspace{-5mm}                 
                 %
                 %
                 \includegraphics[width=0.44\textwidth]{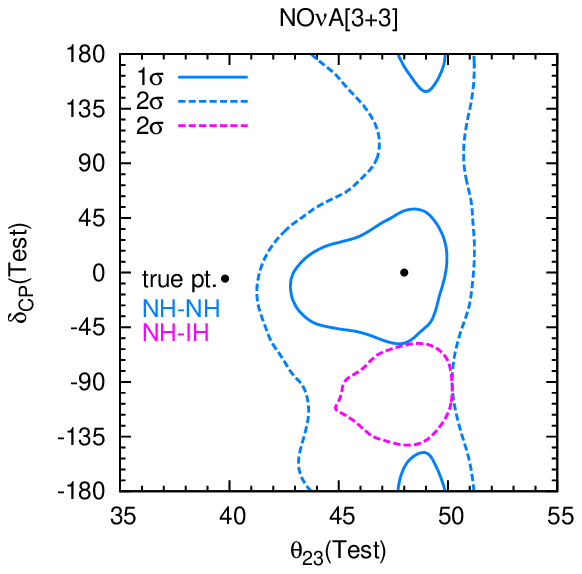}
                 \hspace*{-1.0in}
                \includegraphics[width=0.44\textwidth]{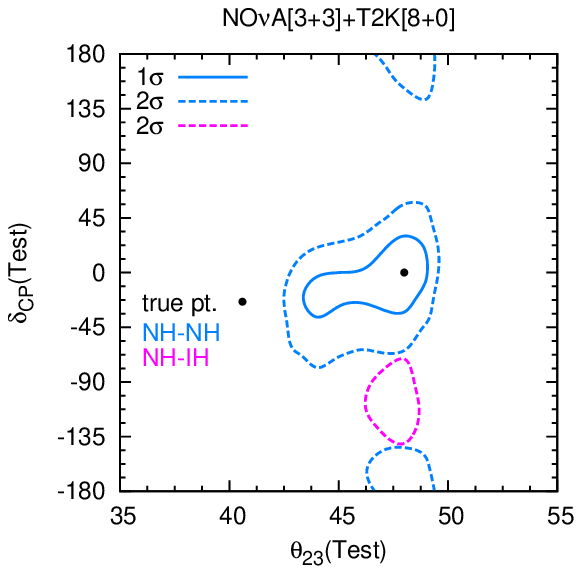}
                 \hspace*{-1.0in}  
                 \includegraphics[width=0.44\textwidth]{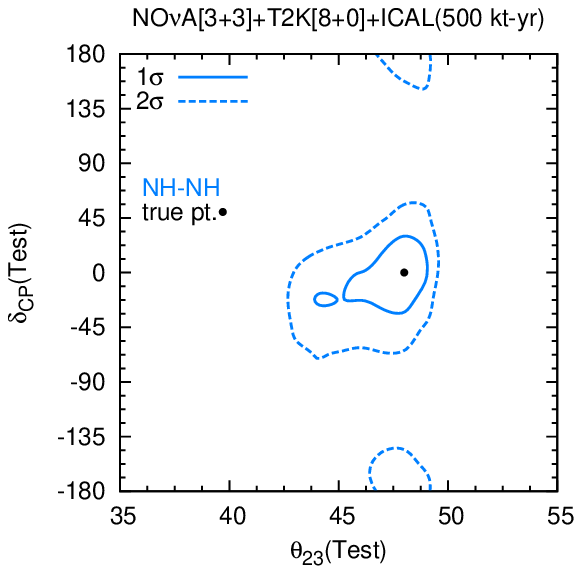} \\
 \vspace{-5mm}                
                  %
                  %
                 \includegraphics[width=0.44\textwidth]{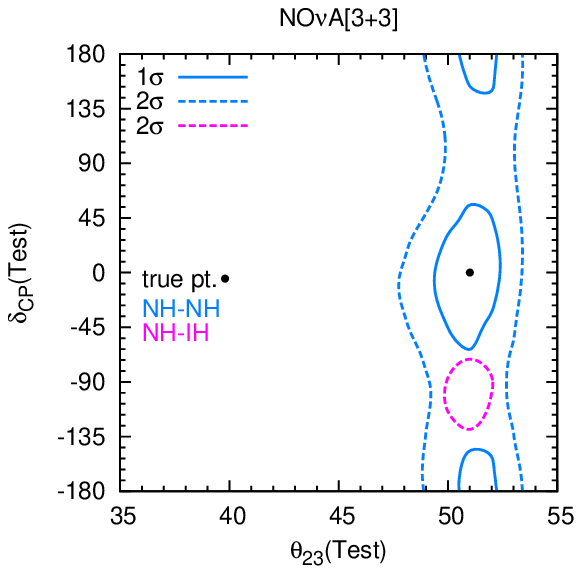}
                 \hspace*{-1.0in}
                \includegraphics[width=0.44\textwidth]{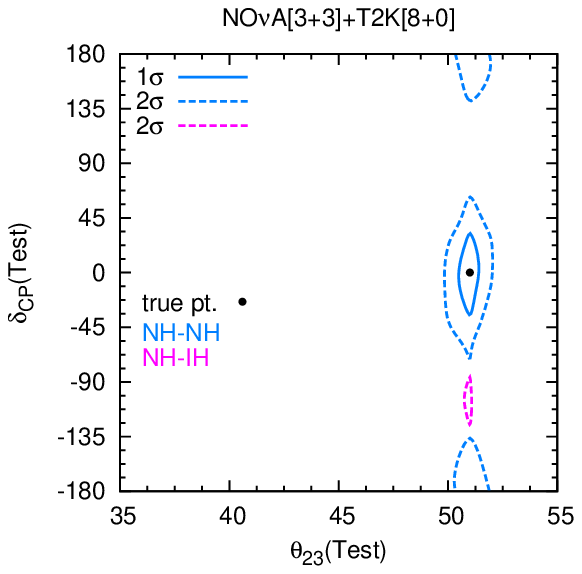}
                 \hspace*{-1.0in}
                 \includegraphics[width=0.44\textwidth]{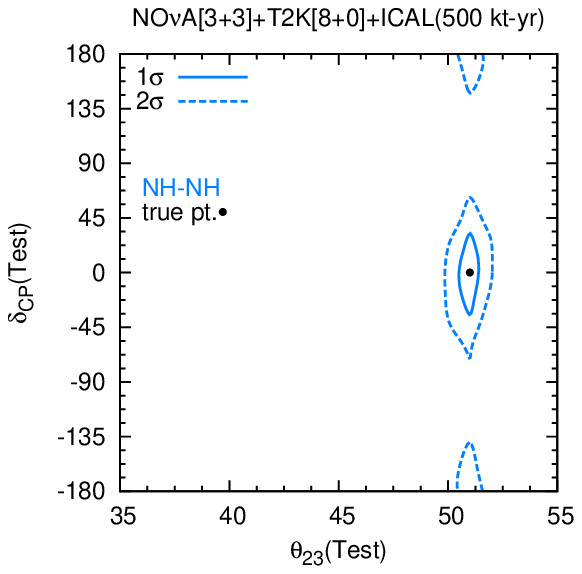} \\

        \end{tabular}
 \vspace{1.3cm}
 \caption{\footnotesize Same as in Fig.\ref{fig:cp_-90} but for true $\dcp= 0^\circ$ .}
\label{fig:cp_0} 
\end{figure}

Our results are presented 
in Figs. ~\ref{fig:cp_-90}, ~\ref{fig:cp_90} and \ref{fig:cp_0} which 
correspond to true $\dcp=-90^\circ$, $90^\circ$ and $0^\circ$
respectively. In each figure the successive columns represent 
\nova[3+3], \nova[3+3]+T2K[8+0], and \nova[3+3]+T2K[8+0]+ICAL respectively.
In these figures the following generic features can be noted:

\begin{itemize}

\item Comparing with the \nova[6+0] panels, in all the \nova[3+3] figures we 
note that the addition of antineutrino information
removes the wrong-octant degenerate regions. 
This also includes the wrong-hierarchy regions occurring with the wrong octant.  For $\theta_{23} = 39 ^\circ$ and $51^\circ$, 
the wrong-octant regions are almost completely removed. 
But for the true $\theta_{23}$ values $42^\circ$ and $48^\circ$, 
both the right-hierarchy and wrong-hierarchy solutions extend to the wrong-octant region.  

\item When T2K data is added to \nova[3+3], it helps in removing these wrong-octant extensions. 
The wrong-hierarchy right octant regions also get significantly reduced in 
size by adding T2K data to \nova[3+3]. This is due to 
the fact that for T2K, these solutions occur at different 
$\dcp$ values than \nova.  Addition of T2K data also improves the precision 
of $\theta_{23}$ and $\dcp$.

\item When ICAL data are added to T2K and \nova, the remaining wrong-hierarchy regions are resolved for all the true values of $\theta_{23}$ 
considered.  The wrong-octant extensions of the right-hierarchy 
solutions are also reduced in size and the precision of $\theta_{23}$ 
improves. 
The combination of T2K+\nova+ICAL can resolve all the degeneracies at 
a $2\sigma$ level for true $\theta_{23} = 39^\circ, 51^\circ$ for all the 
three $\dcp$ values. 
For the $\theta_{23}$ and $\dcp$ combinations of \{$42^\circ$,  $ 0^\circ$\} 
and for {$48^\circ$, $ \pm 90^\circ$, $ 0^\circ $ }, 
there are still allowed regions in the wrong-octant.
 Note that some of the wrong-octant regions that are removed by
the \nova\ antineutrino run could also be removed by the  
ICAL data. 

\end{itemize}
Apart from the above features, the following important points can be observed from the figures: 

\begin{itemize}
\item For $\dcp=-90^\circ$, there are no wrong-hierarchy solutions in 
\nova[3+3], and the addition of T2K  helps in improving the $\theta_{23}$ precision. 
This feature is particularly prominent for $\theta_{23}=42^\circ$ and
$48^\circ$  where T2K data help in removing the wrong-octant extensions for the right
hierarchy solutions. With the addition of  ICAL data, the  wrong-octant 
solution is almost resolved for $\theta_{23} = 42 ^\circ$,  while    
for $\theta_{23} = 48 ^\circ$ the same is resolved at $1 \sigma$.    

\item For $\dcp=-90^\circ$ and $\theta_{23} = 51^\circ$, \nova\ can already 
 resolve all the degeneracies with  6 years of neutrino run only 
 as can be seen in Fig.~\ref{only_nu}. However, the precision of $\theta_{23}$ 
 is worse with \nova[3+3]. This is because splitting the neutrino run into equal neutrino and antineutrino runs reduces the statistics and hence the precision becomes worse. 

\item For $\dcp=90^\circ$, we also see that for $\theta_{23}=48^\circ$, the wrong-hierarchy region in \nova[3+3]+T2K[8+0] is still quite large and this is where ICAL has a remarkable role to play. We see that when ICAL data are added, the large wrong-hierarchy region corresponding to $\theta_{23}=48^\circ$ completely vanishes. 

\item  The small $1\sigma$ wrong hierarchy-wrong octant allowed zone for $\dcp=90^\circ$ 
and $\theta_{23}=42^\circ$ in \nova[3+3] can be identified as the part of
the WH-WO-R$\dcp$ solution of \nova[6+0] by comparing with Fig.~\ref{only_nu}. 

\item For $\dcp=0^\circ$, adding T2K data to \nova[3+3] improves the precision considerably and also removes the wrong-hierarchy solutions to a large extent. 
The precision of $\theta_{23}$ and $\dcp$ around the true solution also 
improves. The enhanced precision due to adding T2K is also responsible for reducing the continuous allowed  regions to discrete degenerate solutions for $\theta_{23}$  values near $45^\circ$. Adding ICAL data removes the remaining wrong-hierarchy regions and 
further help to pinpoint the allowed zones at a $2\sigma$ level.

\item In \nova[3+3], for $\dcp = 0^\circ$ and $\theta_{23} = 39^\circ$, $42^\circ$,
comparing with the corresponding figures 
in Fig.~\ref{only_nu}, we see that the spurious solution appearing at $-150^\circ$ at $1\sigma$  due to the intrinsic degeneracy 
is no longer present with the addition of antineutrino data,
since for the antineutrino probability in \nova\ the intrinsic degeneracy 
is between $0^\circ$ and $+150^\circ$ as discussed earlier. 
Thus the addition of neutrino and antineutrino data solves the intrinsic
degeneracy at $1\sigma$ at both these $\dcp$ values but at 2$\sigma$  
both  $\pm 150^\circ$  remain allowed.
The allowed area near the true value  increases in size because replacing 
half the 
neutrino run with antineutrinos reduces the statistics and hence the 
precision becomes worse.

\end{itemize}

The following additional observations can be made regarding alternative parameter values and running modes:

\begin{itemize}

\item In generating the above plots, we considered T2K running in neutrino mode 
with its full beam power.  
We find that once one includes the antineutrino run from \nova, running T2K 
in the antineutrino mode is no longer necessary for removing spurious 
wrong-octant solutions. Rather, running in the neutrino mode gives enhanced 
statistics and hence better precision. If on the other hand \nova\ runs in
full neutrino mode and the antineutrino component comes from T2K, we have
verified that we get similar results. 

\item We have presented the results for the case of true NH. If the true 
hierarchy is chosen to be IH, one would get a different set of allowed 
regions based on the degeneracies observed in Fig.~\ref{novaprob}. For example, for 
$\dcp = -90^\circ$ and $\theta_{23} = 39^\circ$ for \nova[6+0] in the true IH case, 
apart from the true solution, RH-WO-W$\dcp$ and WH-RO-W$\dcp$ solutions would be 
obtained. This can be predicted from Fig.~\ref{novaprob} (top left panel) by drawing a 
horizontal line from the bottom of the IH-LO band at $\dcp = -90^\circ$, which cuts both the IH-HO and NH-LO bands near $\dcp = 90^\circ$. 

The situation for \nova[3+3] and other combinations would be more complicated 
since the allowed regions and precision for true IH depend not only on the 
probability behavior but also on the statistics of neutrino and antineutrino 
data in the respective experiments.

\item The results are significantly dependent on the true value of $\theta_{13}$, 
chosen here to be $\sin^2 2\theta_{13} = 0.1$. Lower values of 
$\theta_{13}$ (or worse $\theta_{13}$ precision) would lead to 
poorer CP precision and more difficulty in removing the degeneracies. This is because $\dcp$ is coupled with $\theta_{13}$ in the oscillation probability. 

\end{itemize}
\subsection{Distinguishability between $ 0^\circ $ and $ 180^\circ $} 

It will also be interesting to see how far the two CP conserving values $ 0^ \circ $ and $ 180^ \circ $ can be distinguished by the experimental setups considered. 
In this section, we discuss  this issue. The true events are generated for $\dcp=0^\circ$, $\theta_{23}=39^\circ$ and normal hierarchy. In the test spectrum, we consider $\dcp=180^\circ$ and marginalize over $\theta_{13}$. For purposes of comparison we also give the results for test $ \dcp = 90^ \circ $. The results are presented in Table~\ref{tab:0_pi}. We observe that a $2\sigma$ sensitivity in distinguishing between 
 $ \dcp = 0^ \circ $ and $ \dcp = 180^ \circ $ can be achieved by \nova+T2K. Adding 
ICAL data increases the sensitivity further. It is interesting to note that  
for beam based experiments, $ 0^ \circ $ and $ 90^ \circ $ have much larger separation than that between $0^\circ$ and $180^\circ$. But for ICAL, $ 0^ \circ $ and $ 180^ \circ $  are more separated though  the $\chi^2$ values are very small. This is because ICAL itself 
has limited  CP sensitivity due to angular smearing over all directions \cite{mono_ical}. 
Note that for experiments like PINGU the CP sensitivity can be higher, and the $ \chi^{2} $  difference between $ 0^ \circ $ and $ 180^ \circ $ can be appreciable \cite{Razzaque:2014vba}.    
\begin{center}
\begin{table}[H]
\begin{tabular}{| c | c | c | }
\hline
$ \nu + \overline{\nu} $ &Test   $\dcp = 90^ \circ $ & Test $\dcp = 180^ \circ $ \\
\hline  
NO$ \nu $A[3+3] & 6.31 & 2.82  \\
NO$ \nu $A[3+3]+T2K[8+0] & 14.63 & 4.77 \\ 
ICAL & 1.21 & 1.60\\
NO$ \nu $A[3+3]+T2K[8+0]+ICAL & 14.83 & 5.4\\
 \hline 
\end{tabular}
\caption {$ \chi^{2} $ sensitivity for test $ \dcp=90^\circ,~180^\circ $ with true $ \dcp $ = $0^ \circ $, true hierarchy as NH and true $ \theta_{23} $ as $39^ \circ $.}
 \label{tab:0_pi}
\end{table}
\end{center}

\subsection{Precision of $\theta_{23}$ and $\dcp$} 

As stated earlier, the contour plots also give information about the precision 
of $\theta_{23}$ and $\dcp$. In general the presence of degenerate solutions leads to a worse precision  (a larger width of the allowed area) in these parameters.  
For most values of true $\dcp$ and $\theta_{23}$, there is a negligible 
difference between the $\dcp$-precision of \nova[6+0] and \nova[3+3] around the true point. While there is a qualitative advantage to including both neutrinos and antineutrinos because of their different dependences on $\dcp$, this advantage is squandered by the lower cross section of antineutrinos. The precision in these parameters can be quantified using the following formulas: 
\begin{align}
{\mathrm CP~ precision} &=  \dfrac{\dcp^{Max} - \dcp^{Min}}{360^\circ} \times 100 \% \\
{\mathrm \theta_{23}~ precision} &= \dfrac{\theta_{23}^{Max} - \theta_{23}^{Min}}{\theta_{23}^{Max} + \theta_{23}^{Min}} \times 100 \%
\end{align}
In Table ~\ref{tab:precision}, we list the values of the 1$\sigma$ and 
2$\sigma$  precision of $\theta_{23}$ and $\dcp$ using these expressions 
for the case of \nova[3+3] + T2K[8+0] + ICAL. The CP precision is seen to be better for $\dcp =0^\circ$ as compared to $\dcp = \pm 90^\circ$ for a given true value of $\theta_{23}$. This is because in the absence of degeneracies, the precision simply follows from the probability expression, where $dP_{\mu e}/d\dcp$ is smallest at $0^\circ$~\cite{suprabh_t2knova}. On the other hand, for a given value of $\dcp$, the CP precision is seen to become worse with increasing $\theta_{23}$. The $\theta_{23}$ precision is worse near maximal mixing and improves as one moves away. 
\vspace{5mm}
\begin{table}[H]
 \centering
 \begin{tabular}{|c|c|c|c||c|c|c|c|}
 \hline 
 \multicolumn{2}{|c|}{True value} & \multicolumn{2}{|c||}{LO-Precision} & \multicolumn{2}{|c|}{True value} & \multicolumn{2}{|c|}{HO-Precision} \\
 \hline
$\theta_{23}$ & $\delta_{CP}$ & 1$ \sigma $ & 2$ \sigma $ & $\theta_{23}$ & $\delta_{CP}$ & 1$ \sigma $ & 2$ \sigma $    \\ 
  \cline{3-4}   \cline{7-8}
  & &  $ \theta_{23} \hspace{3mm}  \delta_{CP}$  &  $ \theta_{23} \hspace{3mm}  \delta_{CP}$ & & &  $ \theta_{23} \hspace{3mm}  \delta_{CP}$  &  $ \theta_{23} \hspace{3mm}  \delta_{CP}$ \\
 \hline 
          & +90$^\circ $ & 1.02 \hspace{3mm} 26.63  & 2.17 \hspace{3mm} 39.50 &  &+90$ ^\circ $ & 3.15 \hspace{3mm}   28.45 & 7.70 \hspace{3mm} 48.27  \\
$39^\circ$& -90$^\circ $ & 0.89 \hspace{3mm} 34.52& 2.17 \hspace{3mm} 41.52 & $48^\circ$& -90$^\circ$ & 3.15 \hspace{3mm} 30.00& 7.35 \hspace{3mm} 43.22  \\
          & 0$^\circ $ & 0.64 \hspace{3mm} 15.83& 2.04 \hspace{3mm} 28.33 &  & 0$^\circ $ & 4.03 \hspace{3mm} 17.50 & 7.59 \hspace{3mm} 35.80\\
 \hline 
            & +90$^\circ $ & 1.6 \hspace{3mm}   27.00 & 3.32  \hspace{3mm} 38.52 &  & +90$ ^\circ $ & 0.88 \hspace{3mm} 30.32 & 2.16 \hspace{3mm} 43.33\\
$42^\circ$ &-90$^\circ $ & 1.7 \hspace{3mm}  29.77 & 3.31 \hspace{3mm} 41.52 & $51^\circ$& -90$^\circ $& 0.98 \hspace{3mm} 34.48 & 2.16 \hspace{3mm} 45.00\\
           &0$^\circ $ &1.66 \hspace{3mm}   15.83 & 3.08 \hspace{3mm}  29.16  & & 0$^\circ $ & 0.88 \hspace{3mm}  19.16 & 2.16 \hspace{3mm} 37.50\\
 \hline
 \end{tabular}
 \vspace{3mm}
 \caption{\footnotesize{ Percentage precision of parameters $ \theta_{23}$ 
and $\delta_{CP} $ around true value using NO$ \nu $A+T2K+ICAL.}}
 \label{tab:precision}
\end{table}
 \vspace{-1cm}


\hspace{2cm} 
%

\section{Conclusion}

In the era when the value of $\theta_{13}$ was unknown, an eight-fold degeneracy of neutrino oscillation parameters was identified, which included the intrinsic $\theta_{13}$, hierarchy--$ \delta_{CP}$ and octant degeneracies. With the precise measurement of $\theta_{13}$ the intrinsic degeneracy is largely removed and a four-fold degeneracy out of the  original eight -- involving wrong-hierarchy and wrong-octant solutions -- remains  to be solved by the current and upcoming experiments. 
In this paper, we study these degeneracies in detail and propose that the remaining degeneracies  can be studied in the most comprehensive manner by considering the generalized  hierarchy--$\theta_{23}-\dcp$ degeneracy. This degeneracy is continuous for the $P_{\mu e}$ channel. The addition of information on the measurement of $\theta_{23}$ by the $P_{\mu \mu}$ channel gives rise to discrete solutions. These are best visualized by contours in the  test ($\theta_{23} -\dcp$) plane drawn for both right and wrong-hierarchy for different representative values  of true parameters. We show that, depending on whether the wrong-hierarchy and/or wrong-octant solutions occur with right or wrong values of $\dcp$, there can be a total of eight possibilities. We study these possibilities at the probability level for T2K and \nova. At this level, the degeneracy is defined as the equality of the probabilities for different values of parameters. However, at the $\chi^2$ contour level, because of the precision of the experiments, one gets finite allowed regions corresponding to degenerate solutions.  We define a degenerate solution to be one which is distinct from the true solution at the $1\sigma$ level. 

Taking only the neutrino run of  \nova\ as an illustrative example, we identify which of these degenerate solutions actually occur for different representative choices of true parameters. The sample true values that we consider for obtaining the contours are 
$\theta_{23}= 39^\circ,42^\circ,48^\circ$ and $51^\circ$ and $\dcp = \pm 90^\circ,0^\circ$. At the present level of precision, for $\dcp =\ \pm 90^\circ$, the right (wrong) $\dcp$ solutions are those which occur in the same (opposite) half-plane as compared to the true solution. Since $\dcp=0^\circ$ is common to both half-planes, for this case  the right and wrong $\dcp$ solutions  at a particular C.L.  
are inferred from the nature of the contours. The different degenerate solutions obtained are the (i)WH-WO-R$\dcp$,  (ii) RH-WO-W$\dcp$, (iii)WH-RO-R$\dcp$, (iv) RH-RO-W$\dcp$ and (v) WH-WO-W$\dcp$ regions. Although the options i-iii have been noticed in the literature earlier, the option iv which  exists for the same  true $\theta_{23}$
but different $\dcp$ has not been discussed extensively. A probability level discussion was done in Ref.\cite{Minakata:2013eoa}, where it was called $\theta_{23}-\dcp$ degeneracy. However, since it can occur for the same hierarchy and same $\theta_{23}$  we call it \enquote{intrinsic CP degeneracy}. The WH-WO-W$\dcp$ solutions often appear as part of i, given the CP precision of the current experiments. We identify a few points in the true parameter space  where this solution appears as a distinct degenerate solution.  
Note that for a true value of $\theta_{23}$ in the range $48^\circ-51^\circ$ and 
$\dcp$ in the lower half-plane  ($-180^\circ < \dcp < 0^\circ$), the \nova\ neutrino probability being highest cannot be matched by any other combination, and hence no degenerate solutions appear. In this case, only the neutrino run is better as it gives a better precision. In all other cases that we have studied, 3 years of the neutrino and 3 years of the antineutrino run of \nova\ are helpful in removing the wrong-octant solutions i, ii and v  to a large extent.  This also improves the CP precision since the wrong $\dcp$ solutions occurring with the wrong-octant are resolved. Next, we present the results combining \nova[3+3] with T2K[8+0]. It is seen that the synergy between T2K and \nova\ helps in removing the WH-RO-W$\dcp$ solutions for true $\dcp = 0^\circ,90^\circ$. For true $\dcp=-90^\circ$, \nova\ itself is sufficient for removing this degeneracy.
The precision of both parameters also improves when these two sets of information are compounded together. The remaining degenerate solutions at $2\sigma$ can be resolved 
by adding ICAL data. The latter is seen to play an important role in removing the wrong-hierarchy solution for $\theta_{23} = 48^\circ$. In conclusion, we show that the combination of data from different LBL and atmospheric neutrino experiments
can play a crucial role in removing the degeneracies associated with neutrino oscillation parameters, thereby improving the precision of the parameters $\theta_{23}$ and $\dcp$. 
This also paves the way toward an unambiguous determination of these parameters. 
 
 \appendix
 \section{Synergy between appearance and disappearance channel and role of antineutrinos}
\begin{figure}[H]
        \begin{tabular}{lr}
                \includegraphics[width=0.45\textwidth]{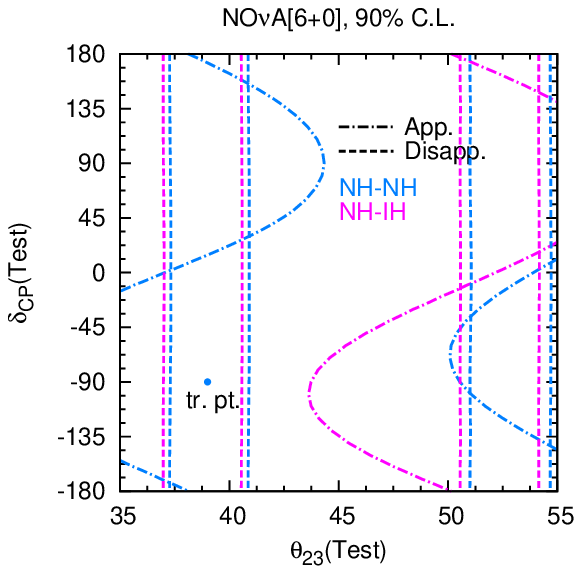}
                & 
                \hspace*{-1.1in}
                 \includegraphics[width=0.45\textwidth]{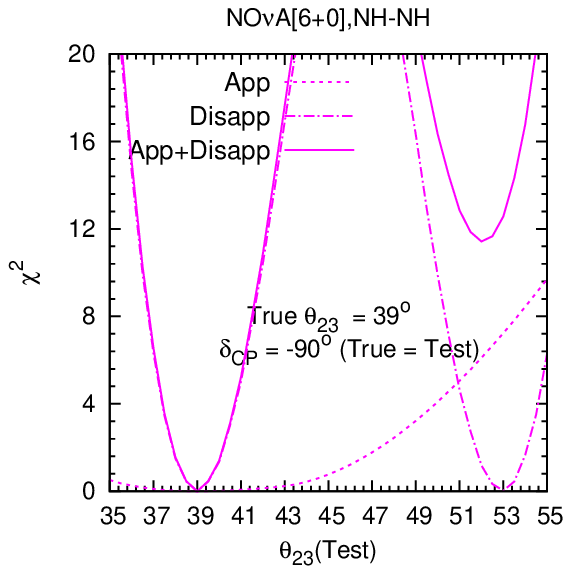}
                 \hspace*{-1.1in}
                 \includegraphics[width=0.45\textwidth]{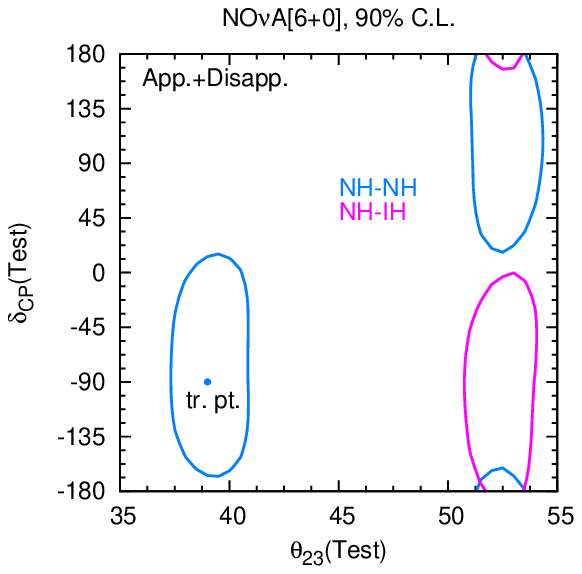}\\
                \includegraphics[width=0.45\textwidth]{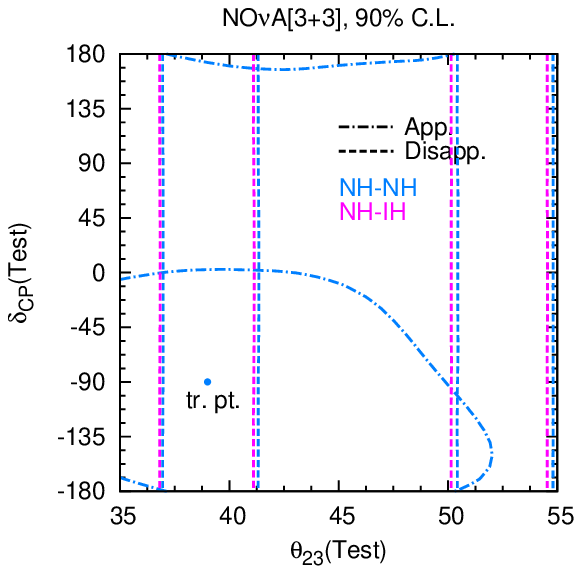}
                & 
                \hspace*{-1.in}
                 \includegraphics[width=0.45\textwidth]{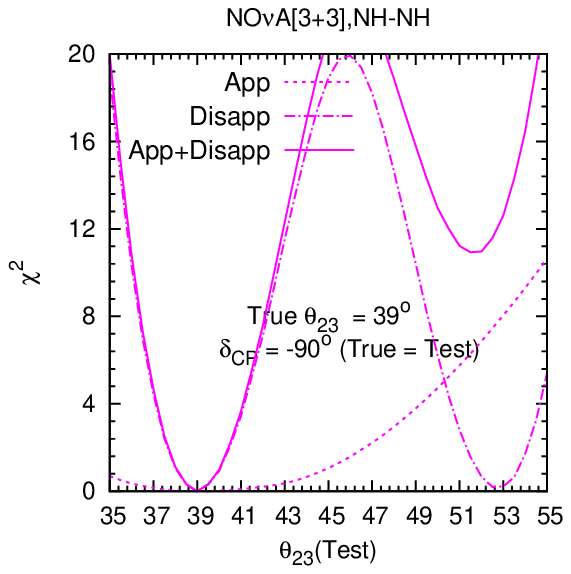}
                 \hspace*{-1.1in}
                 \includegraphics[width=0.45\textwidth]{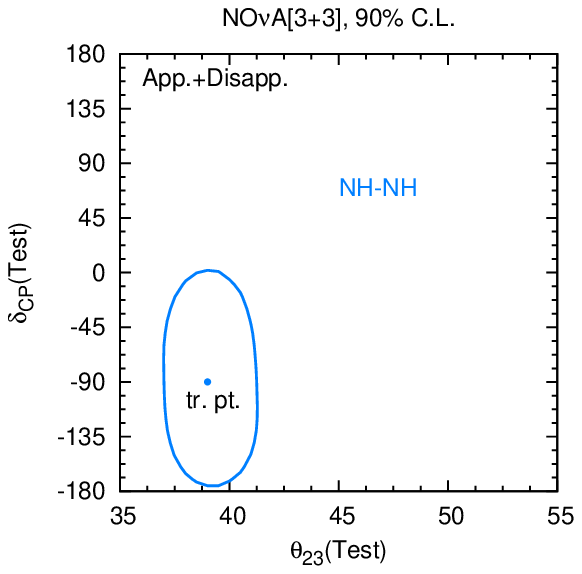}
                      \end{tabular}
\vspace{1.0cm}
\caption{\footnotesize The plots in the upper row are for NO$\nu$A running  only in the neutrino mode i.e. NO$\nu$A[6+0]. Those in the lower row are  for NO$\nu$A running in equal neutrino and antineutrino mode i.e NO$\nu$A[3+3]. To generate these plots, we have assumed  true $ \theta_{23} = 39^{\circ} $, true $ \delta_{CP} = -90^{\circ}$ and true hierarchy = NH, whereas test parameters are marginalized over the range given in Table \ref{param_values}. The plots in the middle panel are generated for a fixed  value of  the test $\dcp $ ($=-90^\circ$). 
}
\label{app_disap}
\end{figure}

 In this Appendix, we discuss the origin of discrete degenerate regions in the test ($\dcp -\theta_{23}$) plane from the combination of  appearance and disappearance channels for \nova. We demonstrate  the role of antineutrinos in resolving the degeneracies. The reference true point  chosen in generating the data is $\dcp=-90^\circ$ and $\theta_{23} = 39^\circ$. In the upper row of Fig. ~\ref{app_disap}, we plot the sensitivity of \nova[6+0]. The serpentine curves in the top-left panel of Fig. ~\ref{app_disap} 
represent the allowed area at 90\% C.L. from only the appearance channel. The  area inside the vertical curves represents the allowed area from only the disappearance channel at the same C.L.  The area between the  blue dotted  (magenta dotted) curves denotes the region obtained for the right (wrong)  hierarchy. For the appearance channel, the allowed region is continuous and no discrete degenerate solutions appear. This can be understood in the following manner. In the neutrino appearance channel, $\dcp=-90^\circ$ corresponds to the maximum value in the probability. As one moves away from $-90^\circ$, the probability decreases and reaches its minimum value  at $+90^\circ$. On the other hand, the  probability increases(decreases) as $\theta_{23}$ increases(decreases). So if we draw an imaginary horizontal line and an imaginary vertical line at the true point, then the allowed region is expected to come along the diagonal of the rectangle obtained by the intersection of these two imaginary lines and the X,Y  axes for $\theta_{23} > 39^\circ$ and $\dcp \leq +90^\circ$. For $\dcp > +90^\circ$, the probability starts to increase, so $\theta_{23}$ has to fall to  keep  the probability same. This explains the serpentine nature of the allowed area. The width of the band corresponds to the  $\theta_{23}$ 
precision of the experiment. For the disappearance channel, the allowed region is in the vicinity of $\theta_{23}$ and $\pi/2 -\theta_{23}$ and parallel to the  $\dcp$ 
axis since the $P_{\mu \mu}$ probability  has a very weak dependence on $\dcp$.  However, the combination of the disappearance and appearance channels gives discrete regions in the parameter space  due to the excellent $\theta_{23}$ precision of the disappearance 
channel near $\theta_{23}=39^\circ$ and $51^\circ$. This helps to exclude the other wrong values of $\theta_{23}$. This is shown in the top right panel of Fig. ~\ref{app_disap}.  Apart from the allowed regions around the true value, one can identify the distinct degenerate solutions corresponding to wrong hierarchy-wrong octant-right $\dcp$  (WH-WO-R$\dcp$) and right hierarchy-wrong octant-wrong $\dcp$ (RH-WO-W$\dcp$) regions.

To show the exact synergy between the appearance and disappearance channels, 
in the middle panel of the top row we plot the $\chi^2$ as a function of $\theta_{23}$(test) for a fixed $\dcp$ value of $-90^\circ$ for the same hierarchy (NH). This figure shows that, though the disappearance channel suffers from the intrinsic octant degeneracy and does not have any octant sensitivity itself {\bf{($\chi^2 \sim 0$)}}, when added to the appearance channel, the channel is responsible for that of $\dcp$. 

Next, we discuss the role of antineutrino runs in \nova. In the bottom row of Fig. ~\ref{app_disap}, we plot the same figures as the top row but for the 3 year neutrino + 3 year antineutrino run. In the bottom left panel, we see that when antineutrino information is added to neutrino data, the allowed region from the appearance channel is significantly reduced. The reason is as follows: as $\dcp$ changes its sign for antineutrinos, the serpentine shape of the allowed region gets flipped with respect to $\dcp$. This excludes the right hierarchy-wrong octant regions of $\dcp \in$ UHP (i.e. RH-WO-W$\dcp$) and the wrong hierarchy-wrong octant regions of $\dcp \in$ LHP (i.e. WH-WO-R$\dcp$). Thus, after adding the antineutrino data only the RH-RO-R$\dcp$ solution remains, as can be seen from the third panel in the bottom row. From the \nova\ antineutrino probability figure in the top right panel of Fig. \ref{novaprob}, it is seen that the probability of the true point cannot be matched by any points in the NH-HO or IH-HO bands. This means that for this true point antineutrinos are free from the  degeneracies that appear with the wrong octant in neutrinos. Thus, the addition of antineutrino information removes the wrong octant solutions of \nova[6+0] that appear in the top right panels of Fig. \ref{app_disap}.

The nature of the disappearance channel contours are seen to remain unaltered but now the allowed area is slightly broader. This is because  of a reduction in the overall statistics due to the smaller cross sections of the antineutrinos. This is also seen in the middle panels where the widths of the $\chi^2$contours increase.

\bibliography{neutosc}

\end{document}